\documentclass[aps,prd,10pt,twocolumn,preprintnumbers,nofootinbib,superscriptaddress,longbibliography]{revtex4-2}

\usepackage{graphicx}
\usepackage{amsmath, amssymb}
\usepackage{hyperref}
\usepackage{graphicx, calc}
\usepackage{comment,xcolor}
\usepackage{booktabs,multirow}
\hypersetup{
  colorlinks  = true,
  urlcolor    = blue,
  linkcolor   = blue,
  citecolor   = red
}
\usepackage[capitalize]{cleveref}
\usepackage{fontawesome}

\newcommand{\Tau}{\mathrm{T}}
\newcommand{\Nu}{\mathrm{N}}
\newcommand{\dd}{\mathrm{d}}

\newcommand{\orcid}[1]{\begingroup
  \hypersetup{hidelinks}\href{https://orcid.org/#1}{\includegraphics[width=10pt]{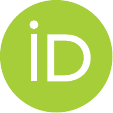}} \endgroup}

\begin{document}

\title{Seafloor Topography Enhances KM3NeT Sensitivity to ANITA-like Events}

\author{Carlos A. Argüelles\orcid{0000-0003-4186-4182}}
\email{carguelles@g.harvard.edu}
\affiliation{Department of Physics \& Laboratory for Particle Physics and Cosmology, Harvard University, Cambridge, MA 02138, USA}

\author{Toni Bertólez-Martínez\orcid{0000-0002-4586-6508}}
\email{bertolezmart@wisc.edu}
\affiliation{Department of Physics, Wisconsin IceCube Particle Astrophysics Center, University of Wisconsin, Madison, Wisconsin 53706, USA}
\affiliation{Departament de Física Quàntica i Astrofísica (FQA), Universitat de Barcelona (UB), c. Martí i Franqués, 1, 08028 Barcelona, Spain}
\affiliation{Institut de Ciències del Cosmos (ICCUB), Universitat de Barcelona (UB), c. Martí i Franqués, 1, 08028 Barcelona, Spain}

\author{Alba Burgos-Mondéjar\orcid{0009-0009-1439-928X}}
\email{ab2910@cam.ac.uk}
\affiliation{Emmanuel College, University of Cambridge, Cambridge CB2 3AP, United Kingdom}

\author{Anne-Katherine Burns\orcid{0000-0002-4391-1731}}
\email{annekatherineburns@icc.ub.edu}
\affiliation{Departament de Física Quàntica i Astrofísica (FQA), Universitat de Barcelona (UB), c. Martí i Franqués, 1, 08028 Barcelona, Spain}
\affiliation{Institut de Ciències del Cosmos (ICCUB), Universitat de Barcelona (UB), c. Martí i Franqués, 1, 08028 Barcelona, Spain}

\author{Jacobo Lopez-Pavon\orcid{0000-0002-9554-5075}}
\email{jlpavon@ific.uv.es}
\affiliation{Instituto de F\'isica Corpuscular (CSIC-Universitat de Val\`encia), Parc Científic UV,C/ Catedr\'atico Jos\'e Beltr\'an 2, E-46980 Paterna (Valencia), Spain}

\author{Jordi Salvado\orcid{0000-0002-7847-2142}}
\email{jsalvado@fqa.ub.edu}
\affiliation{Departament de Física Quàntica i Astrofísica (FQA), Universitat de Barcelona (UB), c. Martí i Franqués, 1, 08028 Barcelona, Spain}
\affiliation{Institut de Ciències del Cosmos (ICCUB), Universitat de Barcelona (UB), c. Martí i Franqués, 1, 08028 Barcelona, Spain}

\date{\today}
\preprint{}

%%%%%%%%%%%%%%%%%%%%%%%%%%%%%%%%%%%%%%%%%%%%%%%%%%%%%%%%%%%%%%%%%%%%%%%%%%%%%%
%%%%%%%%%%%%%%%%%%%%%%%%%%%%%%%%%%%%%%%%%%%%%%%%%%%%%%%%%%%%%%%%%%%%%%%%%%%%%%
\begin{abstract}
In this article, we introduce the concept of \textit{topographic enhancement} in the context of ultra-high-energy neutrino detection by underwater neutrino telescopes.
We demonstrate that the local topography around KM3NeT/ARCA can increase the detection efficiency in scenarios involving long-lived particles by up to a factor of $\sim 3$ due to the presence of an underwater mountain range in the direction of Malta. 
We consider a simplified model-independent approach that parametrizes the new physics able to generate both track-like and cascade-like signals in neutrino telescopes.
When explaining the KM3-230213A event with a diffuse dark flux hypothesis, including its azimuthal direction—in addition to the zenith angle—provides additional constraints on the parameter space.
In this effective model, the observations by KM3NeT and ANITA-IV can be simultaneously explained, but a global tension with the lack of a corresponding detection in IceCube remains at 2.7 sigma.
This work underscores the importance of incorporating topographic effects in the design and optimization of next-generation neutrino telescopes, as is done in the context of mountain-based detectors such as TAMBO. 
We present \href{https://github.com/tbertolez/BSMatUHEdets}{a numerical code~\faGithub} which can be used to easily extend this topographical analysis to other experiments.
\end{abstract}

\maketitle

%%%%%%%%%%%%%%%%%%%%%%%%%%%%%%%%%%%%%%%%%%%%%%%%%%%%%%%%%%%%%%%%%%%%%%%%%%%%%%
\section{Introduction}
\label{sec:intro}
%%%%%%%%%%%%%%%%%%%%%%%%%%%%%%%%%%%%%%%%%%%%%%%%%%%%%%%%%%%%%%%%%%%%%%%%%%%%%%
On February 13, 2023 the Astroparticle Research with Cosmics in the Abyss (ARCA) detector of the KM3NeT project observed an ultra-high-energy (UHE) through-going muon with a reconstructed energy of $120^{+110}_{-60}\, \rm PeV$~\cite{KM3NeT:2025npi}, named KM3-230213A. Reconstruction of the event shows that the muon was traveling nearly horizontally relative to the detector. 
This makes an atmospheric origin for the muon unlikely, and thus the leading hypothesis is that the muon was produced by an UHE neutrino which propagated through Earth and interacted via charge-current interaction~\cite{KM3NeT:2025npi}. 
The reconstructed neutrino energy is $220_{-110}^{+570}\, \rm PeV$, making it potentially the highest-energy neutrino ever detected. 
At this energy, the atmospheric neutrino flux is vanishingly small, indicating that this neutrino likely originated from either astrophysical sources~\cite{KM3NeT:2025bxl} or cosmogenic processes~\cite{KM3NeT:2025vut}.

The mean free path of UHE neutrinos in the Earth is several hundreds of kilometers, so detector sensitivity in the zenithal direction peaks near the horizon, where KM3-230213A was observed. 
However, while the dependence of the column depth on the zenithal angle is always taken into account, this is not the case for the azimuthal angle~\cite{KM3NeT:2020mc}. 
Importantly, ARCA is located $34$~km next to an underwater cliff, making its surroundings in the horizontal plane very asymmetrical~\cite{KM3NeT:2025npi}.
The azimuthal direction of the observed muon, which precisely points towards this cliff, may hold overlooked information on the origin of the event.
Both in present and future detectors, the exploration of the UHE regime requires quantifying the effect of the environment's topography in their sensitivity.

The observation of KM3-230213A is an important step in this exploration, as it has raised many open questions surrounding the nature and origin of the event. 
One of the most intriguing is that no neutrino in this energy range has been observed in the IceCube neutrino telescope, over its more than 12 years of operation and detector volume far exceeding that of ARCA~\cite{IceCubeCollaborationSS:2025jbi}. 
Since the start of its operation, the IceCube collaboration has detected hundreds of astrophysical neutrinos with well understood systematic uncertainties controlled by in-situ measurements~\cite{IceCube:2020wum,Abbasi:2021qfz,IceCube:2021rpz}. 
However, the energy deposition of none of these events exceeds 6 PeV.
This raises a tension between the measurement from KM3NeT and the lack of UHE signals in IceCube, which is between two sigma if the neutrino originated from a transient source, and up to 3.5 sigma for the case of diffuse sources~\cite{KM3NeT:2025npi,KM3NeT:2025ccp,Li:2025tqf, Palmisano:2025abd}.

Furthermore, KM3-230213A is not the first measured UHE event which is compatible with a neutrino origin. 
In particular, in December 2016 the ANITA-IV experiment measured four UHE events, from about 1~degree below the horizon~\cite{ANITA:2020gmv, ANITA:2021xxh}, and shower energies between 0.8 and 3.6 EeV.
\begin{figure}
    \centering
    \includegraphics[width=\linewidth, alt={plain-text
    This plot shows the energy of the neutrino in GeV in the x-axis (between 1e4 and 5e10), and the flux multiplied by energy squared in the y-axis, in units of GeV over centimeter-squared, second and sterorradian, both in log scale. On the left half of the plot, the purple and pink data points (tracks and cascades, respectively) are falling with energy. On the right hand of the plot, the upper bounds of IceCube to UHE neutrinos is shown, it increases with energy. Two data points have a much larger flux: the KM3NeT point, a factor 10 above the sensitivity line at 2e8 GeV, and the ANITA-IV one, a factor 100 above the line, at 1e10 GeV. 
    }]{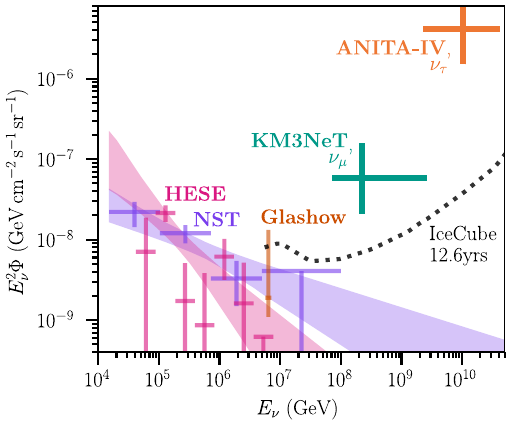}
    \caption{Measurements of high-energy and ultra-high-energy events compatible with a neutrino origin, illustrated through their corresponding differential flux. In orange, the four events from the four flight of ANITA~\cite{ANITA:2020gmv,ANITA:2021xxh}, consistent with a $\nu_\tau$ origin. In teal, the KM3-230213A event from KM3NeT~\cite{KM3NeT:2025npi}, consistent with a $\nu_\mu$ origin. In pink, purple and brown, IceCube measurements for high-energy starting events (HESE)~\cite{IceCube:2020wum}, Northern-Sky Tracks (NST)~\cite{Abbasi:2021qfz} and the Glashow resonance~\cite{IceCube:2021rpz}, respectively. Pink and purple shades represent the power law fits to their respective data, at 68\% confidence level, and their extrapolation to higher energies. Finally, a dotted line represents the IceCube upper limit at $E\gtrsim 10\,\mathrm{PeV}$, at 90\% confidence level~\cite{IceCubeCollaborationSS:2025jbi}. 
    \emph{Under the Standard Model and a diffuse flux hypothesis, both KM3-230213A and ANITA-IV events are in tension with the non-observation of UHE events at IceCube.}}
    \label{fig:SM-fluxes}
\end{figure}
These events look remarkably similar to the event seen by KM3NeT, both in their energy and directionality.
All are Earth-skimming with compatible track lengths, and, as shown in \cref{fig:SM-fluxes}, are similarly in tension with lack of observation in IceCube. 
This similarity could be a hint pointing towards the same physics responsible for both the KM3NeT and ANITA-IV observations. 

While many Beyond the Standard Model (BSM) solutions have been proposed to solve the tension between KM3NeT and IceCube~\cite{Airoldi:2025opo, Sakharov:2025oev, Dev:2025czz, Farzan:2025ydi, Baker:2025cff, He:2025bex, Kohri:2025bsn, Brdar:2025azm, Borah:2025igh, Esteban:2025wbv,Anchordoqui:2025xug,Choi:2025hqt}, 
the tension can only be completely relieved on a transient source hypothesis.
However, the ANITA-IV event can only be reasonably explained under an isotropic diffuse flux hypothesis, where BSM scenarios can relieve the tension between ANITA-IV and IceCube~\cite{Bertolez-Martinez:2023scp}.
Following a model independent approach, in this article we parametrize a class of BSM models which can produce both the KM3NeT and ANITA-IV signals, and study the consistency of these observations in the global UHE landscape under a diffuse flux hypothesis.
To this end, we present the first analysis, to our knowledge, that accounts for the topography surrounding an underwater UHE neutrino telescope, namely, the ARCA detector.
A topographic analysis is not only mandatory for incorporating all the information from KM3-230213A, but the additional freedom of our model independent approach allows us to understand under which conditions the sensitivity of ARCA can be boosted--an effect which we call \textit{topographic enhancement}--, and how azimuthal information can be leveraged to constrain BSM models in UHE detectors.

Our paper is organized as follows. In \Cref{sec:formalism} we describe our BSM model independent approach and assumptions considered in the analysis. In \Cref{sec:KM3NeT-angular} we study the azimuthal dependence of KM3NeT sensitivity and discuss the information that can be extracted in our framework considering only KM3-230213A. In \Cref{sec:anita-iv} we show that the scenario under consideration can simultaneously explain the events observed by both ANITA-IV and KM3NeT. In \Cref{sec:icecube} we discuss the compatibility of these measurements with IceCube in our framework. Finally, we conclude in \Cref{sec:conclusions}. 

%%%%%%%%%%%%%%%%%%%%%%%%%%%%%%%%%%%%%%%%%%%%%%%%%%%%%%%%%%%%%%%%%%%%%%%%%%%%%%
\section{Formalism}
\label{sec:formalism}
%%%%%%%%%%%%%%%%%%%%%%%%%%%%%%%%%%%%%%%%%%%%%%%%%%%%%%%%%%%%%%%%%%%%%%%%%%%%%%

In this Section, we introduce a model independent effective parameterization of a class of BSM models that can produce both cascade- and track-like events. Our intention is to create a model-independent setup with the minimal assumptions and parameters required to describe propagation and detection of generic UHE particles, minimally extending the effective model introduced in~\cite{Bertolez-Martinez:2023scp} to account for both cascade and track signals. 

\begin{figure}
    \centering
    \includegraphics[width=\linewidth, alt={plain-text 
    This illustration is split vertically in two panels. In both panels, an incoming N particle is arriving from the left and interacting with matter with cross-section sigma, after which a long-lived particle T is produced. In the top panel, the T particle decays in the air and produces a cascade with branching ratio $Br(T\to casc)$, which is measurable in the flying ANITA-IV. In the bottom panel, the T particle decays into muons, which can then produce a track and arrive to KM3NeT underwater.}]{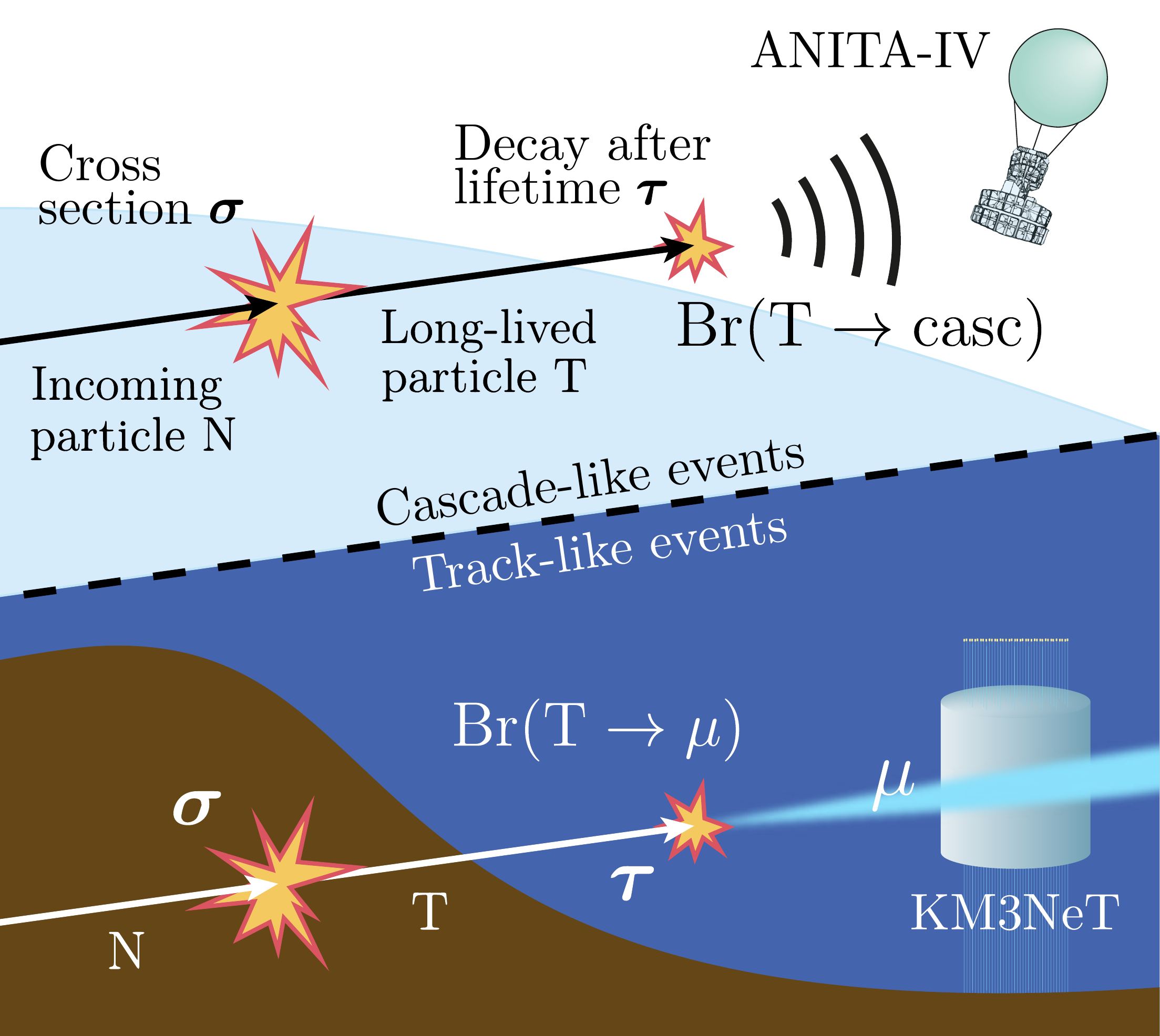}
    \caption{A BSM origin for the ANITA-IV and KM3-230213A events. An incoming flux of dark particles produces, with cross section $\boldsymbol{\sigma}$, secondary long-lived particles which decay after a laboratory lifetime $\boldsymbol{\tau}$. These particles can decay into muons and produce track-like topologies as in KM3-230213A (bottom), or into other SM particles able to produce electromagnetic cascades as in ANITA-IV (top). We will conservatively assume $\mathrm{Br}(\Tau\to\mathrm{casc})=1-\mathrm{Br}(\Tau\to\mu)$. \emph{While $\sigma$ and $\tau$ control the propagation, the branching ratios control the track-to-cascade ratio.}}
    \label{fig:BSM-process}
\end{figure}

\Cref{fig:BSM-process} shows how cascades and tracks can be produced. In this class of BSM models, an incoming flux of dark particles, denoted N, interacts with matter with a cross-section $\sigma$. In this interaction, a long-lived particle (LLP), denoted T, is produced and then decays after a lab-frame mean lifetime $\tau$. If T decays into a muon, with a branching ratio $\mathrm{Br}(\Tau\to\mu)$, this muon will produce a track-like event if it arrives at the detector. If, instead, T decays into another hadronic or electromagnetic particle, it will produce a cascade-like event. We will assume that T does not decay only into invisible particles, i.e., $\mathrm{Br}(\Tau\to\mu)+\mathrm{Br}(\Tau\to\mathrm{casc})=1$. 
Therefore, only one extra effective parameter, $\mathrm{Br}(\Tau\to\mu)$, will be added with respect to the simplified scenario considered in~\cite{Bertolez-Martinez:2023scp}.   

Particular BSM realizations that could be described by this parameterization have been explored in~\cite{Bertolez-Martinez:2023scp,Dev:2025czz,Farzan:2025ydi,Heurtier:2019rkz}.
Importantly, our parameterization can approximate the SM in two particular scenarios. On the one hand, $\sigma = \sigma_{\rm SM}$, $\tau = 0$ (and $\mathrm{Br}(\Tau\to\mu)=1$) mimics a $\nu_\mu$ directly producing a muon track. On the other hand, $\sigma = \sigma_{\rm SM}$ and $\tau = \tau_{\rm SM}$ mimics a $\nu_\tau$ producing a tau lepton, with mean lifetime $\tau_{\rm SM}$, which can then decay into muons. In the SM, this happens with ${\rm Br}(\Tau\to\mu)=0.17$, but here we treat this branching ratio as a free parameter. In the following, 
\begin{equation}\label{eq:SMxs}
\begin{split}
    \sigma_\mathrm{SM} &\simeq 2\times 10^{-32} \, \mathrm{cm}^2 \, , \\ 
	\tau_\mathrm{SM} &\simeq 2\times 10^{-4} \, \mathrm{s} \, ,
\end{split}
\end{equation}
are defined as the SM neutrino charged-current cross-section at $10\, \rm EeV$, and the tau lepton lab-frame lifetime at $1\, \rm EeV$.
We refer to these points in the $(\sigma,\tau)$ space as the $\nu_\mu$-like and $\nu_\tau$-like scenarios, respectively. 
Finally, note that this parameterization does not capture all possible BSM scenarios. Models not based on long-lived particles have been proposed to explain both ANITA events~\cite{Esteban:2019hcm} and KM3-230213A~\cite{Sakharov:2025oev, Airoldi:2025opo, Baker:2025cff, He:2025bex, Kohri:2025bsn, Brdar:2025azm, Borah:2025igh, Esteban:2025wbv,Anchordoqui:2025xug}.

While the observation of one event in KM3NeT in one year of operation makes a transient origin for KM3-230213A a valid solution~\cite{Li:2025tqf, Farzan:2025ydi, Dev:2025czz, Yuan:2025zwe, Zhang:2025abk,Neronov:2025jfj,Das:2025vqd,deOliveira:2025ufx,Sakharov:2025oev,Palmisano:2025abd}, 
this is not the case for ANITA-IV, which observed four events from different directions in only one month of operation. 
Thus, a joint explanation of KM3NeT, ANITA-IV and IceCube requires assuming a common diffuse flux origin. 
We introduce a flux of dark particles with unknown normalization, $\Phi$. This normalization being unknown is a key aspect of our analysis, which will determine the possible constraints derived from the data.

As a first assumption, we restrict the incoming flux to a narrow energy window consistent with the cascade energies observed in the ANITA-IV events, specifically between $0.8$ and $3.6\, \mathrm{EeV}$~\cite{ANITA:2020gmv}. This EeV
flux can still produce $120\, \mathrm{PeV}$ through-going muons due to muon energy losses, which keeps the setup minimal yet effective. 
While monochromatic spectra are characteristic of BSM scenarios such as dark matter decay~\cite{Kohri:2025bsn,Borah:2025igh} or primordial black hole evaporation~\cite{Airoldi:2025opo,Choi:2025hqt}, given the low statistics of UHE telescopes a monochromatic line is in practice indistinguishable from a broader spectrum confined to a narrow energy range, such as the peak region of a broken power law
, e.g., ~\cite{Rodrigues:2020pli,Bergman2020,Valera:2022ylt}. 
Since ANITA's exposure to radio showers is in its plateau region at these energies, growing only slowly with energy, 
the event rate from such a spectrum is dominated by a narrow range around $E_\nu \sim 1$\,EeV.
In any case, a power-law flux extending to lower energies would overproduce PeV events in IceCube, beyond the sensitivity range of ANITA-IV. In the following sections we will show that even under the favorable assumption of a monochromatic flux, the tension with IceCube remains above $2\sigma$.

As a second assumption, we neglect T and N energy losses, and assume that T can be absorbed by matter with the same cross-section $\sigma$ as in production, as expected from time-reversal invariance of the production process. Additionally, we assume fixed energy transfers from N to T and from T to muons, respectively. In \Cref{sec:absorption} we study the impact of relaxing this assumptions in our results, showing that our main conclusions remain robust.

In our framework, cascade-like events can be produced in three different ways: in the scattering of an N or T particle, or in the decay of T into a cascade. 
Since we assume T to be uncharged, track-like events can only be produced by the decay of T into a muon. If the T particle decays inside the detector, the event will be a starting track and if this occurs outside the detector but close enough to it, the event will be a through-going muon.

Up to the branching ratio factors that will be introduced below, the number of events per unit of solid angle generated by decays and scatterings inside the detector is given by
\begin{align}
\label{eq:T_events}
\frac{\mathrm{d}N^\Tau_\mathrm{dec}}{\mathrm{d}\Omega} &=  \frac{\Phi}{4\pi} \, P^\Tau_\mathrm{exit} \, P^\Tau_\mathrm{decay} \, A_\mathrm{g}\,  \varepsilon \, \Delta t \, , \\ 
\frac{\mathrm{d}N^{\Nu(\Tau)}_{\rm scat}}{\mathrm{d}\Omega} &= \frac{\Phi}{4\pi} \, P^{\Nu(\Tau)}_\mathrm{exit} \, \sigma \, N_\mathrm{targets} \, \varepsilon \, \Delta t\, , 
\label{eq:N_events}
\end{align}
respectively. Here, $\Phi$ is the all-sky flux normalization, $A_{\rm g}$ the geometrical area of the detector (see \cref{sec:muon-energy-losses} for details on its calibration), $\varepsilon$ the detector efficiency, $\Delta t$ the observation time and $N_{\rm targets}$ the number of targets in the instrumented volume of the detector. $P^{\rm N(T)}_{\rm exit}$ is the probability that an N (T) particle arrives at the instrumented volume (see Appendix~\ref{sec:probabilities}). Finally, $P^\Tau_{\rm decay} = 1-e^{-d/c\tau}$ is the probability that T decays inside the instrumented volume, of depth $d$. In ANITA-IV the detection efficiency depends on the altitude of the decay point~\cite{PierreAuger:2025hvl}. However, this dependence is not publicly available for the angular range relevant to our analysis. In \cref{app:decay-altitude} we implement a simplified extrapolation of the altitude dependence, and find that the ANITA-IV best-fit $\tau$ lifetime shifts
to a value shorter by a factor of $\sim 0.6$, yet our overall conclusions remain valid.

Given our set of assumptions, in our simplified framework the signals from decay and scatterings inside the instrumented volume are monochromatic, but muon energy losses spread track-like signals to low energies for through-going muons. Our study matches the official KM3NeT analysis~\cite{KM3NeT:2025npi} and thus accounts for all muons with energy larger than $10\, \rm PeV$, thus
\begin{equation}
    \frac{\mathrm{d}N_\mu}{\mathrm{d}\Omega} = 
    \frac{\Phi}{4\pi}A_{\mathrm{g}}\,\varepsilon
    \int_{10\, \mathrm{PeV}}^{E_0} \dd E^\mu_{\rm fin} \,
    P_\mu(E_{\rm fin}^\mu)\, ,
\end{equation}
where $P_\mu(E_{\mathrm{fin}}^\mu)$ is the probability that a muon is produced and arrives to the detector with energy $E^\mu_{\rm fin}$. \Cref{sec:subPeV-muons} shows that considering muons with $E^\mu_{\rm fin} <10\, \rm PeV$ does not modify our conclusions.

Then, the three possible event topologies are given by
\begin{align} \nonumber
     &\frac{\mathrm{d}N_\mathrm{casc}}{\mathrm{d}\Omega}  =  \frac{\mathrm{d}N^\Tau_\mathrm{dec}}{\mathrm{d}\Omega}[1-\mathrm{Br(\Tau\to\mu)}] + \frac{\mathrm{d}N^{\Tau}_{\rm scat}}{\mathrm{d}\Omega} + \frac{\mathrm{d}N^{\Nu}_{\rm scat}}{\mathrm{d}\Omega}\, , \\ 
     &\frac{\mathrm{d}N_\mathrm{start}}{\mathrm{d}\Omega}  =  \frac{\mathrm{d}N^\Tau_\mathrm{dec}}{\mathrm{d}\Omega}\mathrm{Br(\Tau\to\mu)}\, , \\ 
    &\frac{\mathrm{d}N_\mathrm{through}}{\mathrm{d}\Omega}  =  \frac{\mathrm{d}N_\mu}{\mathrm{d}\Omega}\mathrm{Br(\Tau\to\mu)}\, , \nonumber
\end{align}
for cascades, starting tracks and through-going muons, respectively. Analytic expressions for the key quantities introduced here are further detailed in \cref{app:detailed-computations}. In the next section we show that the number of events depends not only on the zenithal angle but also on the azimuthal angle.

%%%%%%%%%%%%%%%%%%%%%%%%%%%%%%%%%%%%%%%%%%%%%%%%%%%%%%%%%%%%%%%%%%%%%%%%%%%%%%
\section{Topographic enhancement at KM3NeT}
\label{sec:KM3NeT-angular}
%%%%%%%%%%%%%%%%%%%%%%%%%%%%%%%%%%%%%%%%%%%%%%%%%%%%%%%%%%%%%%%%%%%%%%%%%%%%%%

\begin{figure}
    \centering
    \includegraphics[width=\linewidth, alt={plain-text
    This is a polar plot, where north is oriented to the top. The radial distance to the center shows the effective area to muons of ARCA, up to 5000 meters squared. Two curves are shown. The first one is for a cross-section equal to the SM, and zero lifetime. This has a value of roughly 3300 meters squared from -40º to 175º (north-east), and a value of 2500 meters squared between 180º and 315º (south west), smooth transition in the middle. The second one is for a cross-section equal to the SM, and lifetime equal to the tau lepton at 1EeV. The area is 3100 meters squared between -45º and 175º (north east), and 5000 between 225º and 290º (south west), with a smooth transition in the middle. The curves are overlaid over a topographic map around the ARCA detector, showing a cliff in the west/south of the detector, and water to the north/east. Sicily and Malta are seen to the north-west and west, respectively.}]{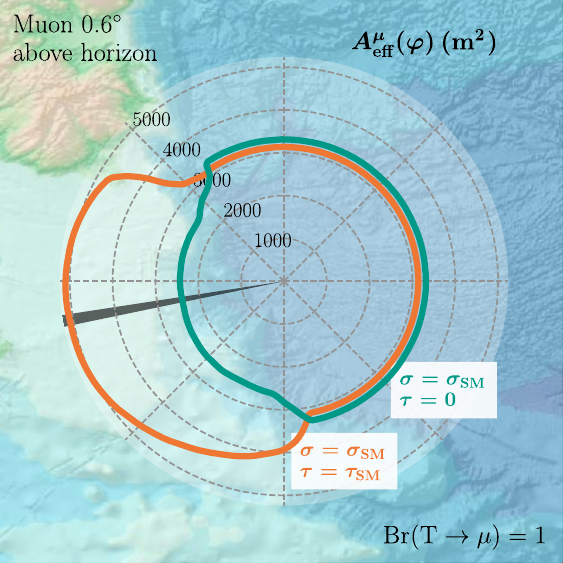 }
    \caption{Azimuthal distribution of the ARCA effective area to muons under different   $(\sigma,\tau)$ parameters, overlaid on the topography surrounding the ARCA detector~\cite{GEBCO_2024_Grid}. The chosen elevation angle, $0.6^\circ$ above horizon, and the azimuthal angle shown in a grey band coincide with the reconstructed direction of KM3-230213A. In teal, ARCA is slightly more sensitive to $\nu_\mu$-like events from water-dominated trajectories, as explained in the main text. Analogously, in orange, ARCA is more sensitive to $\nu_\tau$-like events from rock-dominated trajectories. Here, $\mathrm{Br}(\Tau\to\mu)=1$. \emph{The incoming direction of KM3-230213A may favor a BSM interpretation with an intermediate long-lived particle.}}
    \label{fig:azimuthal-plot}
\end{figure}

In this Section, we quantify the topographical asymmetry of the surrounding area around the ARCA detector and show how it can affect the angular distribution of the expected number of events, both in BSM and in the $\nu_\mu$/$\nu_\tau$-like scenarios. Further, we show that the observation of the KM3-230213A event constrains the parameter space, favoring a long-lived particle explanation.

As reported by the KM3NeT collaboration~\cite{KM3NeT:2025npi}, KM3-230213A is most likely a muon coming from $0.6^\circ$ above the horizon and at an azimuth of $259.8^\circ$ (north at $0^\circ$, increasing clockwise), with an uncertainty of $1.5^\circ$ at 68\% confidence level.
This Earth-skimming direction is expected in the SM, since at $E\gtrsim \mathcal{O}(100)\,\mathrm{PeV}$ the mean free path of neutrinos in Earth is $\mathcal{O}(100)\, \mathrm{km}$. To constrain cross-sections comparable with the SM at these energies, it is essential to have a detailed knowledge of the horizontal plane around the detector.
To such purpose, we use high-resolution bathymetric data from GEBCO~\cite{GEBCO_2024_Grid}, which provides elevation information for both the seabed and adjacent land surfaces. We join this topography with a PREM model for Earth's interior~\cite{DziewonskiAnderson1981PREM}, and compute the width and density of all layers along any given trajectory towards the detector. 

\begin{figure*}[ht]
    \centering
    \includegraphics[width=\textwidth,alt ={plain-text
    This is a collection of four plots. In each of them, the effective area to muons is shown in the different angular directions from the ARCA detector (zenithal between -5º and 5º, azimuthal between 0º and 360º), and each plot has a different set of (sigma, tau) parameters. In the top left, the nu_mu-like scenario is shown, and the effective area peaks to positive zenithal angles, and prefers the water over the cliff. In the bottom left, the nu_tau-like scenario wants horizontal events, and prefers the cliff. In the top right, with a lifetime 1000 times larger, the effective area is smaller and prefers slightly upgoing events. This is also the case for the bottom right, which has a cross-section 100 times smaller.}]{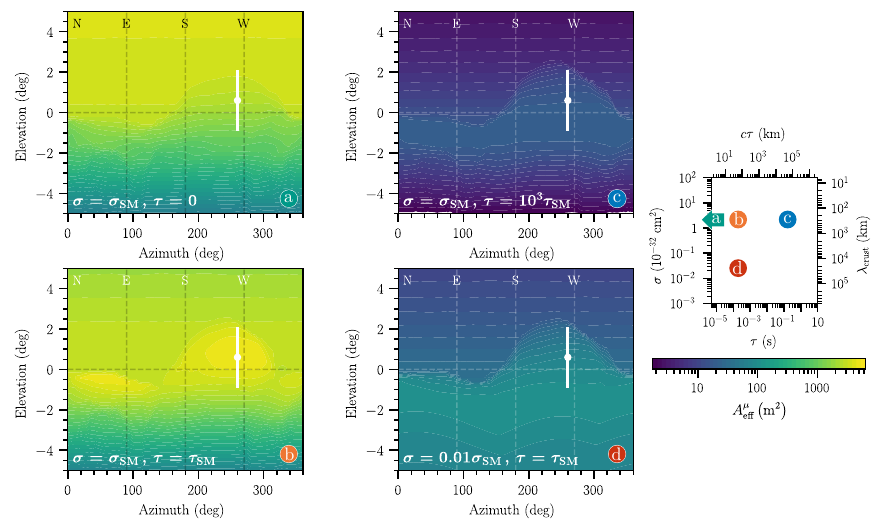} 
    \caption{Angular dependence of the muon effective area $A_{\rm eff}^\mu$ (as defined in the main text) for the different points of the $(\sigma,\tau)$ parameter space as chosen in the rightmost plot. A common color bar for all plots shows $A_{\rm eff}^\mu$ around the horizontal plane, and the white point indicates the reconstructed direction of KM3-230213A within 68\% confidence region. Long lifetimes and small cross sections favor up-going trajectories as shown in the right panels, while the SM-like scenarios shown in the left panels favor Earth-skimming trajectories. \emph{The angular distribution of ARCA events has the potential to constrain their BSM origin.}}
    \label{fig:angular-distributions}
\end{figure*}

\Cref{fig:azimuthal-plot} shows the asymmetry of the surroundings of the ARCA detector close to the horizon due to the underwater cliff, located about 34~km from the detector as measured along the most likely direction of KM3-230213A~\cite{KM3NeT:2025npi}.
This uneven topography translates directly into a direction-dependent effective area to muons, which is shown here as the radial distance to the center, for a muon $0.6^\circ$ above horizon.
This effective area $A_{\rm eff}^\mu$ is defined as $\dd N_\mu/\dd \Omega = \Phi/(4\pi)\Delta t\, A_{\rm eff}^\mu$, with $\Delta t$ the total time of observation, as if ${\rm Br}(\Tau\to\mu)=1$.
The computation of effective areas is detailed in \cref{app:detailed-computations}. Teal and orange curves show $A_{\rm eff}^\mu$ for a $\nu_\mu$-like and $\nu_\tau$-like event, respectively. For a $\nu_\mu$-like scenario, with $\tau=0$, the primary vertex happens in water and close to the detector. Therefore, north/east trajectories, with only water and thus a smaller flux absorption, are slightly favored.
On the contrary, for a $\nu_\tau$-like scenario, i.e., with an intermediate long-lived particle, the primary interaction does not necessarily happen close to the detector.
South/west trajectories, where production can happen in the rock of the cliff, are then favored over water-only trajectories.
Notably, given identical branching ratios, the direction of KM3-230213A matches better the latter scenario, by a factor $\sim 2$ over the conventional
$\nu_\mu$-like explanation. We define \textit{topographic enhancement} as the increase of the effective area in the direction of the rock, which can be computed as the ratio of $A_{\rm eff}^\mu$ between the SW and NE directions.
In the $\nu_\tau$-like scenario this ratio is $\sim 1.6$, but in some regions of the parameter space it can increase up to $\sim 3$.

\cref{fig:angular-distributions} shows the dependence of $A_{\rm eff}^\mu$ on the azimuthal and zenithal directions, for different illustrative choices of the $(\sigma,\tau)$ parameters. 
While trajectories with less elevation traverse longer distances, the azimuthal effect of the cliff on the available column depth is significant. 
The $\nu_\mu$-like scenario shown in the top left panel, $(\sigma=\sigma_{\rm SM},\tau=0)$, prefers downward-going trajectories where N particles have been less absorbed.
In these directions, however, large backgrounds from atmospheric muons will hinder signal detection and azimuthal cuts might need to be considered so that sufficient overburden is present. 
The bottom left panel corresponds to  the $\nu_\tau$-like scenario, $(\sigma=\sigma_{\rm SM},\, \tau=\tau_{\rm SM})$, in which the angular distribution peaks to the west, in the direction of the cliff next to ARCA, that coincides with the observed event. 
Longer lifetimes or smaller cross sections predict events further below the horizon, as shown in the right panels of the figure, with longer track lengths available. 
In conclusion, topographical analyses of detector surroundings are essential to extract the full potential of the information provided by the angular distribution of the observed events. To such purpose, we provide \href{https://github.com/tbertolez/BSMatUHEdets}{a numerical code~\faGithub} which can be used to easily extend this topographical analysis to other experiments.

The parameter space can be further constrained by considering the type of signal. 
Since ARCA observed a through-going muon, models predicting a large ratio of through-going muons to starting tracks and cascades are favored. This rules out scenarios with large $\sigma$ or $\tau$. 
However, the official collaboration analysis has not searched for cascade-like signals~\cite{KM3NeT:2025npi}. In the SM, ARCA's sensitivity to cascades is $\sim$5 times smaller than to tracks, and cascade-like events are unlikely~\cite{KM3Net:2016zxf}. While in our framework cascades can dominate over tracks, in our analysis we still use that KM3NeT has only looked for track-like signals. \Cref{sec:test-statistics} shows how an additional search for cascades would constrain large $\tau$ more strongly.
Since cascades are not searched for, in our analysis of KM3NeT alone, $\Phi$ and $\mathrm{Br}(\Tau\to\mu)$ are completely degenerate as $\Phi\,{\rm Br}(\Tau\to\mu)$.  

\begin{figure}[ht]
    \centering
    \includegraphics[width=\linewidth,alt={plain-text
    This plot shows the (sigma, tau) parameter space, with the lifetime in the x-axis between 1e-6 and 10 s, and the cross-section between 1e-35 and 1e-30 centimeters squared. Green filled contours show the 1 sigma and 2 sigma preferred regions for these parameters. The one sigma is a closed contour around the middle of the plot, with cross-sections larger than 1e-33 and 1e-31, and lifetimes between 1e-4 and 0.1. The two sigma is an open contour which expand mainly to the left and to the bottom, including arbitrarily small cross sections and lifetimes. In the background of this plot, the flux normalisation is shown, which is minimum in the top left corner (small lifetime, large cross section) with 0.1 over kilometer squared and day, and increases to the bottom right.}]{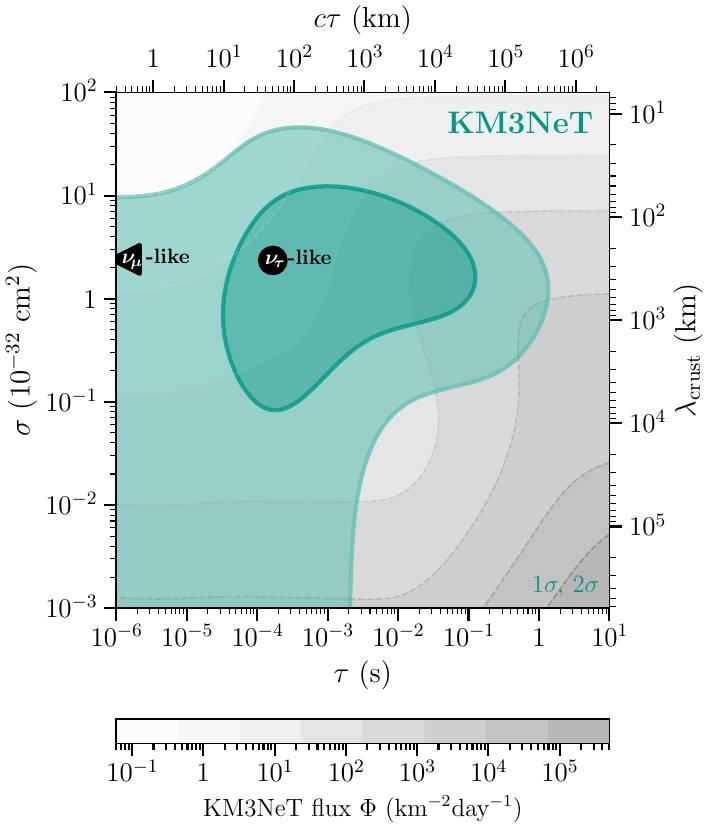}
    \caption{Allowed $(\sigma,\tau)$ parameter space from KM3-230213A under the diffuse flux hypothesis, accounting for its angular direction and track-like topology, at 1 and 2 sigma confidence levels. On the background, the required flux normalisation to produce one event in the detector's observation time, with $\mathrm{Br}(\Tau\to\mu)=1$. Additional axes show the mean free path of N particles in Earth's crust, $\lambda_{\mathrm{crust}}$, and decay length, $c\tau$; and the $\nu_\mu$-like and $\nu_\tau$-like scenarios are shown as a black arrow and circle, respectively. \textit{KM3-230213A alone can already constrain to 1 sigma the BSM regions that can explain it.}}
    \label{fig:KM3NeT-test-statistic}
\end{figure}

\Cref{fig:KM3NeT-test-statistic} shows that KM3-230213A can only be explained in some regions of the parameter space, using the test statistic defined in~\cref{sec:test-statistics}. The arrow and circle show the $\nu_\mu$-like and $\nu_\tau$-like scenarios, respectively, where our framework approximates the SM. Here, the unknown parameter $\Phi\,{\rm Br}(\Tau\to\mu)$ is profiled over, i.e., set to produce one muon event in 335 days of ARCA observation. As a consequence, the constraints on $(\sigma,\tau)$ come only from the angular direction of the event and the fact that the muon is not produced inside the detector.

The test statistic ---detailed in \cref{sec:test-statistics}-- discards cross-sections much larger than $\sigma_{\rm SM}$, since events would peak much more above the horizon; while small cross sections are discarded because events should peak much below the horizon.
\cref{fig:KM3NeT-test-statistic} shows that scenarios involving a LLP with lab decay length of $\mathcal{O}(10-10^4)\, {\rm km}$ are slightly preferred. 
The $\nu_\tau$-like scenario is allowed within one sigma, while the $\nu_\mu$-like scenario is disfavored at 1.5 sigma. Note, however, that in the SM the effective area to $\nu_\tau$ must be corrected by the branching ratio of the tau lepton into muons; for a given flux, this makes the SM $\nu_\tau$ event rate smaller than the $\nu_\mu$ one.
Finally, \cref{fig:KM3NeT-test-statistic} also shows the required flux normalization with ${\rm Br}(\Tau\to \mu)=1$. For the best-fit parameters, $\sigma = 2.2\times 10^{-32}\, \mathrm{cm}^2$ and  $\tau= 1.0\times 10^{-3}\, \mathrm{s}$, $\Phi=20\, \mathrm{km}^{-2}\mathrm{day}^{-1}$ is an order of magnitude larger than the flux of UHE cosmic rays above $1\, \rm EeV$~\cite{PierreAuger:2021hun}. \Cref{app:muon-transfer} shows that a different treatment of the energy transfer to muons would only modify the 
required flux by 20\% at most.
Below, we show that this explanation can also accommodate the observation of the ANITA-IV events. 

%%%%%%%%%%%%%%%%%%%%%%%%%%%%%%%%%%%%%%%%%%%%%%%%%%%%%%%%%%%%%%%%%%%%%%%%%%%%%%
\section{Consistency with ANITA-IV}
\label{sec:anita-iv}
%%%%%%%%%%%%%%%%%%%%%%%%%%%%%%%%%%%%%%%%%%%%%%%%%%%%%%%%%%%%%%%%%%%%%%%%%%%%%%

Intriguingly, the part of the parameter space preferred by the observation of KM3-230213A, as shown in Fig.~\ref{fig:KM3NeT-test-statistic}, is compatible with the preferred region obtained in the analysis of the ANITA-IV events performed in~\cite{Bertolez-Martinez:2023scp}. Nevertheless, the framework used in~\cite{Bertolez-Martinez:2023scp} was slightly different from the one considered here, since track signals were not accounted for. 
In this Section, we show the compatibility between KM3-230213A and ANITA-IV in this extended model, where the free parameter $\mathrm{Br}(\Tau\to\mu)$ allows both track and cascade signals.

As illustrated in \cref{fig:BSM-process}, $\mathrm{Br}(\Tau\to\mu)$ drives the cascade-to-track ratio. 
Since ANITA-IV events are cascades and KM3-230213A is a track, the branching ratio and flux normalization can match the number of events measured both in ANITA-IV and KM3NeT, for any choice of $(\sigma,\tau)$ (see details in \cref{sec:test-statistics}). 
Then, it is directionality and the fact that KM3-230213A is a through-going-muon which can further constrain $(\sigma,\tau)$.

\begin{figure}[t]
    \centering
    \includegraphics[width=\linewidth, alt={plain-text
    This plot shows the same as the previous one, but has added the contours from ANITA and the joint fit between KM3NeT and ANITA. The ANITA one sigma contour is also closed, between 2e-33 and 2 centimeters squared, and 2e-6 and 3 seconds, with the two sigma contour being open to much smaller cross sections. The joint fit has a one sigma contour between 1e-4 and 0.1 seconds, and 5e-33 and 3e-32 centimeters squared. the two sigma contour is also closed, between 2e-6 and 0.3 seconds, and between 1e-33 and 4e-32 centimeters squared.
    }]{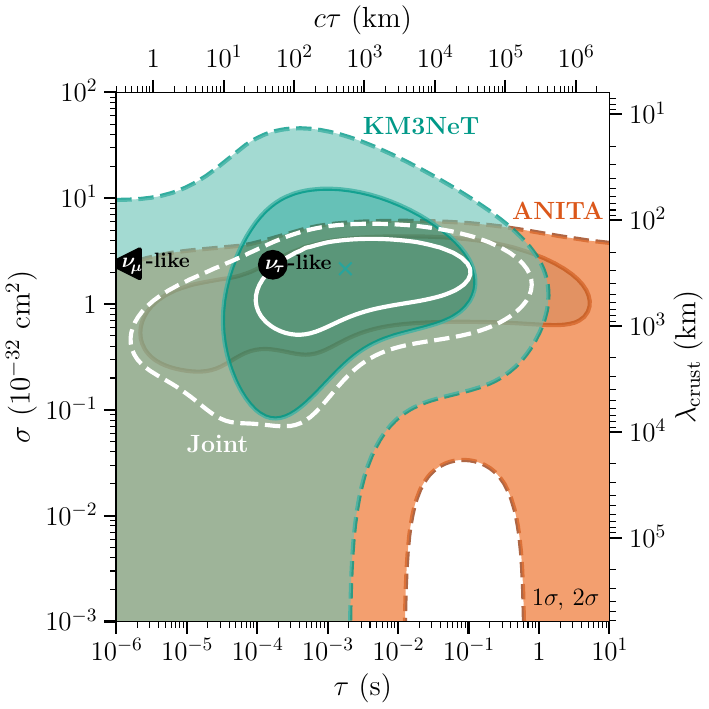}
    \caption{Allowed $(\sigma,\tau)$ parameter space from the joint fit of KM3NeT and ANITA-IV under the diffuse flux hypothesis, after profiling for the flux normalisation and branching ratio. Teal contours show one (solid) and two (dotted) sigma allowed regions for KM3NeT alone, and orange contours similarly for ANITA-IV alone. A cyan dashed region shows the contours for their joint fit. \emph{KM3-230213A and ANITA-IV are compatible with each other, and point to a similar region of the parameter space.}}
    \label{fig:ANITAK3-test-statistic}
\end{figure}

\Cref{fig:ANITAK3-test-statistic} shows the preferred region of the $(\sigma,\tau)$ parameter space for KM3NeT, ANITA-IV and its combination after the profiling over $\Phi$ and $\mathrm{Br}(\Tau\to\mu)$. 
Their significant overlap leads to a combined closed contour at two sigma. Both KM3-230213A and ANITA-IV events are Earth-skimming with a compatible track length, which favors cross sections comparable or slightly smaller than the SM expectation. 
Note that very long lifetimes are allowed by ANITA-IV but, as shown in the previous section, this would generate too many tracks much below the horizon in KM3NeT. Similarly, ANITA-IV alone is compatible with shorter lifetimes than KM3NeT because such short lifetimes would lead to more down-going muons, specially away from the cliff, that have not been observed. 
Finally, the pure $\nu_\mu$-like scenario can hardly explain the ANITA-IV events, which requires $\tau\neq 0$. The best-fit point of our KM3NeT+ANITA joint analysis is
\begin{equation}
    \begin{split}
        \sigma_{\rm BF} &\simeq 2.1\times 10^{-32}\, \mathrm{cm}^2\, , \\
        \tau_{\rm BF} &\simeq 1.8\times 10^{-3}\, \mathrm{s}\, , \\
        \Phi_{\rm BF} &\simeq 34\, \mathrm{km}^2\,\mathrm{day}^{-1}\, , \\
        {\rm Br}(\Tau\to\mu)_{\rm BF} &\simeq 0.92 \, .
    \end{split}
    \label{eq:bestfit}
\end{equation}
Due to KM3NeT's lower exposure, the flux required to explain KM3-230213A is larger than for the ANITA-IV events. This is the reason why in the combined analysis the required branching ratio is close to one and the joint flux closer to the KM3NeT best fit. However, the analyses presented here, specially that of ANITA-IV, are a simplified estimate and not a comprehensive experimental analysis. Our quantitative results should be understood as such. For instance, as shown in \Cref{app:exp-variation}, we expect that the under- or overestimation of both effective areas induced by a relaxation of our assumptions 
leads to a variation of the  
required flux $\Phi$ by 30\% and of the best fit of $\rm Br(\Tau\to\mu)$ between 0.77 and 0.97. 

Finally, the left panel of \cref{fig:flux-events} shows the flux normalization and the predicted number of events for the best-fit value of the combined analysis given in \cref{eq:bestfit}.
On the one hand, the flux required to simultaneously explain ANITA-IV and KM3NeT observations is slightly larger than the one obtained in the official KM3NeT analysis indicated by the gray band in the figure~\cite{KM3NeT:2025npi}.
On the other hand, we observe that the number of events of both experiments is perfectly fitted. However, the fit predicts $\mathcal{O}(100)$ events in IceCube, as expected due to its larger exposure. 
Below we will analyze and quantify how this tension can be slightly relaxed in our effective model.

%%%%%%%%%%%%%%%%%%%%%%%%%%%%%%%%%%%%%%%%%%%%%%%%%%%%%%%%%%%%%%%%%%%%%%%%%%%%%%
\section{Compatibility with IceCube}
\label{sec:icecube}
%%%%%%%%%%%%%%%%%%%%%%%%%%%%%%%%%%%%%%%%%%%%%%%%%%%%%%%%%%%%%%%%%%%%%%%%%%%%%%

\begin{figure*}[ht]
    \centering
    \includegraphics[width=\textwidth, alt ={plain-text
    This plot has 6 panels divided in two rows and three columns, with each column being showing the best fit of a different joint analysis (from left to right, ANITA+KM3NeT, KM3NeT+IceCube, and global fit). The upper row shows the best fit fluxes (in inverse kilometer squared and day), while the lower row shows the number of events. In the ANITA+KM3NeT fit, the ANITA only flux is 2, with the KM3NeT one and the joint fit are around 30. A gray band between 1 and 10 shows the reported flux by the KM3NeT official collaboration. The number of events predicts exactly 4 events in ANITA (as measured), 1 in KM3NeT (as measured) and around 100 in IceCube (around 80 of them are cascades, and the rest are mainly through-going muons). In the KM3NeT+IceCube, the KM3NeT-only fit is 10, and the joint is 0.3. No events are expected in ANITA, around 0.02 in KM3NeT and around 1 in IceXCube, half of them being cascades and half through-going muons. In the global analysis, ANITA-only requires a flux of 3, KM3NeT-only of 30, and the joint of 2. Around 2 events are expected in ANITA, 0.02 in KM3NeT (0.01 are through-going muons) and 2 in IceCube (1.5 of them being cascades).}]{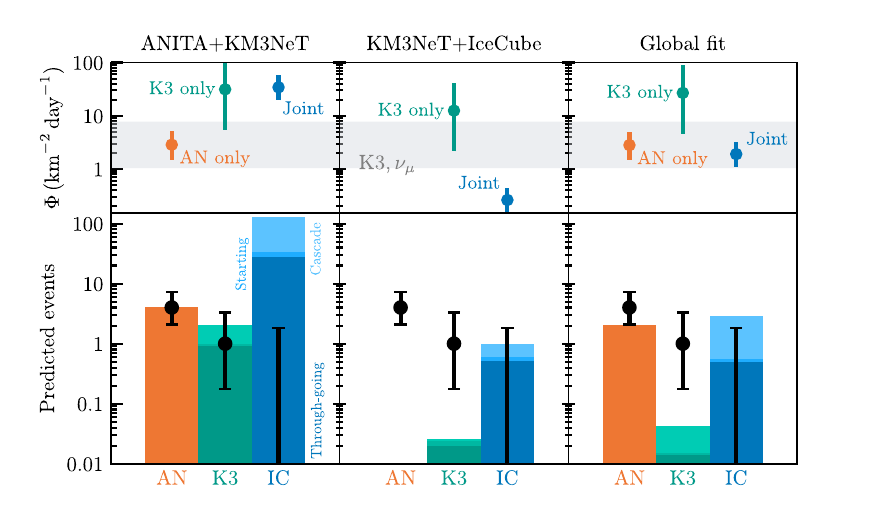} 
    \caption{Flux normalization (upper row) and predicted number of events (lower row) at ANITA-IV (orange), KM3NeT (teal) and IceCube (blue), for the best fits of the three different joint analysis as indicated in the top of the figure. The gray band indicates the flux necessary to explain KM3-230213A within the SM~\cite{KM3NeT:2025npi}. While blue shows the flux normalization from the corresponding joint fit, the orange (teal) dot shows the normalization needed to explain the events observed by ANITA-IV (KM3NeT) alone for this point of the parameter space. In the lower row, orange, teal and blue bars show the predicted number of events at ANITA-IV, KM3NeT and IceCube, respectively. For KM3NeT and IceCube, contributions from different topologies are shown: from darker to lighter colours, through-going muons, starting tracks and cascades. \textit{In all possible fits, too many events are expected in IceCube.}}
    \label{fig:flux-events}
\end{figure*}

The IceCube observatory in the South Pole is the leading experiment in neutrino astronomy, with an instantaneous exposure larger than KM3NeT and a time of observation much larger than ANITA-IV. However, as of today IceCube has not yet identified any signal compatible with KM3NeT and ANITA-IV observations~\cite{IceCubeCollaborationSS:2025jbi}. In this Section, we show that a diffuse-flux global interpretation of the ANITA-IV and KM3NeT events is challenging even given the freedom of the model independent approach considered in this work. The results of this section are summarized in \cref{tab:tensions}.

Thanks to their different geometry, ANITA-IV and IceCube can be made compatible with a cross section slightly smaller than $\sigma_{\rm SM}$ and a lifetime longer than $\tau_{\rm SM}$~\cite{Bertolez-Martinez:2023scp}.
Therefore, as a first step we analyze the compatibility between KM3-230213A and the absence of IceCube events as shown in the middle panel of \cref{fig:flux-events}. This has two challenges. First, KM3NeT and IceCube have analogous experimental efficiencies, so the most important difference between them is the instantaneous observation window, which has been used in other studies to resolve this tension considering transient sources~\cite{Li:2025tqf, Farzan:2025ydi, Dev:2025czz, Yuan:2025zwe, Zhang:2025abk, Neronov:2025jfj, Das:2025vqd,deOliveira:2025ufx,Sakharov:2025oev,Palmisano:2025abd}. However, a diffuse flux interpretation does not benefit from this difference. Second, as of today KM3NeT has a significantly smaller exposure than ANITA-IV, and thus the flux required for KM3-230213A is larger and predicts more events in IceCube than in the analysis of~\cite{Bertolez-Martinez:2023scp}.

The KM3NeT and IceCube joint fit, as defined in \cref{sec:test-statistics}, moves the KM3NeT alone best-fit point obtained in Sec.~\ref{sec:KM3NeT-angular} to $\sigma = 1.1\times 10^{-32}\, \mathrm{cm}^2$ and $\tau = 2.4\times 10^{-4}\, \mathrm{s}$, while keeping ${\rm Br}(\Tau\to\mu)=1$ to increase the rate of track-like events. 
Compared to the KM3NeT-only, the joint analysis prefers smaller cross sections and shorter lifetimes which reduce the cascade-to-track ratio by reducing the detectability of the cascades originated in the primary vertex and shortening the decay and production of the muon.
The non-observation of IceCube also pushes down $\Phi$ to reduce the expected number of events. 
As shown in the central panel of \cref{fig:flux-events}, the best-fit point predicts $\sim 0.02$ through-going muons in KM3NeT and $\sim 1$ events in IceCube. Under this null hypothesis, the observed data has a p-value of $p\sim 0.026$, which corresponds to a $2.2$ sigma tension. That is, our effective model can only slightly relax the tension compared to those presented in the literature for the SM~\cite{KM3NeT:2025npi,Li:2025tqf,KM3NeT:2025ccp,Palmisano:2025abd}. 

The inclusion of ANITA-IV in the global fit, shown in the right panel of \cref{fig:flux-events}, requires ${\rm Br}(\Tau\to\mu)<1$. This increases the cascade-to-track ratio, which favors IceCube's sensitivity compared to that of KM3NeT, and potentially increases their tension.
The geometry of ANITA-IV favors longer lifetimes, which moves the best-fit point to $\sigma = 0.80\times 10^{-32}\, \mathrm{cm}^2$, $\tau = 1.0\times 10^{-3}\, \mathrm{s}$, $\Phi = 1.9\, \mathrm{km}^{-2}\,\mathrm{day}^{-1}$ and ${\rm Br}(\Tau\to\mu)=0.21$. 
This predicts that $\sim 0.014$ through-going muons are expected in KM3NeT, $\sim 2.0$ cascades in ANITA, and $\sim 2.8$ events in IceCube, lowering the total p-value to $p\sim 0.018$. While this is a 
2.7 sigma tension that still remains, the flexibility of our effective model allows for a 
improvement over the SM explanation, which predicts at least 10 times more events in IceCube~\cite{Romero-Wolf:2018zxt}. In \Cref{app:exp-variation}, we show that relaxing the main assumptions in our model independent approach can worsen this tension up to 3.3 sigma. However, the preferred regions of the parameter space do not change qualitatively. 

\begin{table*}
\centering
\begin{tabular}{llc}
\toprule
\textbf{Joint fit} & \textbf{Framework} & \textbf{Tension} \\
\midrule
\multirow{2}{*}{ANITA-IV $+$ IceCube} & SM, diffuse flux~\cite{Romero-Wolf:2018zxt} & $\gg 3\,\sigma$ \\
 & BSM, diffuse flux~\cite{Bertolez-Martinez:2023scp} & $\sim 0.1\,\sigma$ \\ \midrule
\multirow{2}{*}{KM3NeT $+$ IceCube} & SM, diffuse flux~\cite{KM3NeT:2025npi,KM3NeT:2025ccp,Li:2025tqf, Palmisano:2025abd} & 2.5--3.5\,$\sigma$ \\
 & BSM, this work & 2.2\,$\sigma$ \\ \midrule
\multirow{3}{*}{ANITA-IV $+$ KM3NeT $+$ IceCube}  & SM, diffuse flux & $\gg 3\,\sigma$ \\
& BSM, this work & 2.7\,$\sigma$ \\
 & BSM, this work, relaxed assumptions (App.~B) & 2.3--3.3\,$\sigma$ \\
\bottomrule
\end{tabular}
\caption{Summary of the tensions between the different UHE observations under a diffuse flux hypothesis, in the SM and in the effective framework of this work. While the tension of IceCube and KM3NeT with ANITA-IV has not been properly quantified in the SM, the detection of 4 events in ANITA-IV predicts the observation of at least $\mathcal{O}(10^3)$ events in IceCube~\cite{Romero-Wolf:2018zxt}. \textit{In all scenarios, our effective framework can provide a better fit, but tensions always remain above $\sim 2$ sigma.}}
\label{tab:tensions}
\end{table*}

%%%%%%%%%%%%%%%%%%%%%%%%%%%%%%%%%%%%%%%%%%%%%%%%%%%%%%%%%%%%%%%%%%%%%%%%%%%%%%
\section{Conclusions and Outlook}
\label{sec:conclusions}
%%%%%%%%%%%%%%%%%%%%%%%%%%%%%%%%%%%%%%%%%%%%%%%%%%%%%%%%%%%%%%%%%%%%%%%%%%%%%%

The detection of KM3-230213A, with a horizontal direction crossing an underwater cliff, has gained a lot of attention as it could mean the first detection of a neutrino with energy above 100 PeV. This is an energy regime where the yet undiscovered cosmogenic neutrino component is expected to dominate. 
In this article, we have incorporated the detailed topography around the ARCA detector and quantified the azimuthal asymmetry of its effective area. 
With this improved detector description, in \cref{sec:formalism} we have introduced a model independent approach able to parameterize a generic class of BSM models involving long-lived particles which can potentially accommodate this observation and approximately encompasses a $\nu_\mu$ and $\nu_\tau$-like origin. 
Namely, a primary particle N scatters in matter with cross section $\sigma$ and produces a long-lived particle T which decays after a mean lifetime $\tau$, with branching ratio ${\rm Br}(\Tau\to\mu)$ to muons and $1-{\rm Br}(\Tau\to\mu)$ to cascades (see \cref{fig:BSM-process}). 

Using the topographical information around ARCA, in \cref{sec:KM3NeT-angular} we find that the effective area to muons can vary azimuthally by a factor $\sim 3$ in certain points of the $(\sigma,\tau)$ parameter space; we call this efficiency anisotropy: \textit{topographic enhancement}. 
Therefore, we translate the angular distribution and type of event of KM3-230213A into bounds on the allowed $(\sigma,\tau)$. 
We observe that the type of event and direction of KM3-230213A (towards the cliff west of the ARCA detector) prefers $\tau \neq 0$, contrary to the SM $\nu_\mu$ interpretation. 
In particular, we find a one sigma preferred region at $\sigma\in (0.1,10)\times\sigma_{\rm SM}$ and $\tau \in (1,10^3)\times\tau_{\mathrm{SM}}$. 

In \cref{sec:anita-iv} we study KM3-230213A and the ANITA-IV events together and find that they point towards the same preferred region. 
We conclude that both experiments are in good agreement under a diffuse BSM flux interpretation. This is at the expense of large fluxes, much above the sensitivity limits of IceCube. 
We quantify the tension between KM3NeT and IceCube in \cref{sec:icecube}. Even though KM3NeT and IceCube have a similar geometry but KM3NeT much lower exposure, the tension can be reduced to 2.2 sigma in our framework. 
Furthermore, when including also the anomalous ANITA-IV events in the fit, the effective model still can accommodate all observations with a tension of 2.7 sigma. This best-fit region corresponds to cross sections comparable to or slightly below the SM expectation, $\sigma \simeq (0.6\text{--}2)\times 10^{-32}\,\mathrm{cm}^2$, lab-frame lifetimes $\tau \simeq 10^{-3}\,$s, and fluxes $\Phi \simeq \mathcal{O}(1\text{--}10)\,\mathrm{km}^{-2}\,\mathrm{day}^{-1}$, an order of magnitude above the UHE cosmic-ray flux at these energies. We emphasize that even when $(\sigma,\tau)$ in our parameter space coincide with the SM values, the corresponding particles (i.e., $\nu_\mu$-like and $\nu_\tau$-like) do not correspond to SM $\nu_\mu$ and $\nu_\tau$, which require specific branching ratios, and include tau electromagnetic interactions and energy losses.

As shown in \Cref{app:exp-variation}, relaxing the most critical  assumptions in our model independent approach, the ANITA-IV and KM3NeT observations remain highly compatible and there are no significant modifications on the $(\sigma,\tau)$ preferred parameter space. However, this can lead to a larger global tension among ANITA-IV, KM3NeT and IceCube up to 3.3 sigma. The analyses presented here, specially that of ANITA-IV, are a simplified estimate and would be improved by a comprehensive experimental analysis. While it has not been quantitatively carried out, similar results are expected for Auger, since Auger and IceCube share a similar exposure to UHE events.

In summary, while our diffuse BSM flux interpretation improves the compatibility with observations compared to the SM, even under the most favorable assumptions the tension remains above $\sim 2\sigma$. This highlights the difficulty of explaining the KM3NeT event in a global context without invoking a transient origin or an upward fluctuation of the flux.
The observation or non-observation of further events in neutrino telescopes will shed light into the origin and nature of KM3-230213A and the anomalous ANITA-IV events. This is specially exciting in light of the future generation of UHE experiments, starting with PUEO \cite{PUEO:2020bnn}, and following with IceCube-Gen2 \cite{IceCube-Gen2:2020qha}, TAMBO \cite{TAMBO:2025jio}, GRAND \cite{GRAND:2018iaj}, TRIDENT \cite{TRIDENT:2022hql} or P-ONE \cite{P-ONE:2020ljt}.  
Most importantly, given the short mean free path of UHE neutrinos, all these experiments must not overlook the topographical surroundings around their respective detectors.  Local geography, such as underwater cliffs or mountains, can absorb or boost expected signals. 

%%%%%%%%%%%%%%%%%%%%%%%%%%%%%%%%%%%%%%%%%%%%%%%%%%%%%%%%%%%%%%%%%%%%%%%%%%%%%%
%%%%%%%%%%%%%%%%%%%%%%%%%%%%%%%%%%%%%%%%%%%%%%%%%%%%%%%%%%%%%%%%%%%%%%%%%%%%%%
\begin{acknowledgments}
We would like to thank Alfonso Garcia-Soto for illuminating discussions. This work has been supported by the Spanish MCIN/AEI/10.13039/501100011033 grants PID2022-126224NB-C21 and by the European Union’s Horizon 2020 research and innovation program under the Marie Skłodowska-Curie grants HORIZON-MSCA-2021-SE-01/101086085-ASYMMETRY and H2020-MSCA-ITN-2019/860881-HIDDeN. TB is supported by the Spanish grant PRE2020-091896. JLP acknowledges financial support from the Spanish Research Agency (Agencia Estatal de Investigaci\'on) through grant IFIC Centro de Excelencia Severo Ochoa No CEX2023-001292-S, grants PID2023-148162NB-C21 and PID2022-137268NA-C55, and grant CNS2022-136013 funded by MICIU/AEI/10.13039/501100011033 and by “European Union NextGenerationEU/PRTR'', and from the MCIU with funding from the European Union NextGenerationEU (PRTR-C17.I01) and Generalitat Valenciana (ASFAE/2022/020). TB, A-KB and JS acknowledge support from the ``Unit of Excellence Maria de Maeztu 2020-2023'' award to the ICC-UB CEX2019-000918-M.
C.A.A. are supported by the Faculty of Arts and Sciences of Harvard University, the National Science Foundation (NSF), the NSF AI Institute for Artificial Intelligence and Fundamental Interactions (IAIFI), the Canadian Institute for Advanced Research (CIFAR), the David and Lucile Packard Foundation, and the Research Corporation for Science Advancement.
\end{acknowledgments}

%%%%%%%%%%%%%%%%%%%%%%%%%%%%%%%%%%%%%%%%%%%%%%%%%%%%%%%%%%%%%%%%%%%%%%%%%%%%%%
%%%%%%%%%%%%%%%%%%%%%%%%%%%%%%%%%%%%%%%%%%%%%%%%%%%%%%%%%%%%%%%%%%%%%%%%%%%%%%

\bibliographystyle{JHEP.bst}
\bibliography{biblio}

\clearpage
\onecolumngrid
\appendix
\renewcommand\thefigure{\thesection\arabic{figure}}

\setcounter{figure}{0}

%%%%%%%%%%%%%%%%%%%%%%%%%%%%%%%%%%%%%%%%%%%%%%%%%%%%%%%%%%%%%%%%%%%%%%%%%%%%%%
\section{Detailed computations}
\label{app:detailed-computations}
%%%%%%%%%%%%%%%%%%%%%%%%%%%%%%%%%%%%%%%%%%%%%%%%%%%%%%%%%%%%%%%%%%%%%%%%%%%%%%

\subsection{Flux and number of events}
We parametrize the incoming flux by a normalization constant, the energy spectrum and the angular distribution,
\begin{equation}
	\Phi(\Omega, E_\Nu) = \Phi_0\, f_E(E_\Nu)\, f_\Omega(\Omega)\, ,
\end{equation}where $E_\Nu$ is the energy of the incoming particle, $\Omega$ the solid angle ($\mathrm{d}\Omega = \sin\theta\, \mathrm{d}\theta\, \mathrm{d}\varphi$), and $\Phi_0$ the flux per unit area, time, solid angle, and energy. For a diffuse flux, $f_\Omega(\Omega)=(4\pi)^{-1}$. The expected number of events per unit solid angle and final energy is
\begin{equation}
     \frac{\mathrm{d}N(\theta)}{\mathrm{d}\Omega\,\mathrm{d}E_{\rm fin}} = \int_{E_{\text{min}}}^{E_\text{max}}\mathrm{d}E_\Nu\, \Phi_0\, \Delta t\, f_E(E_\Nu)\, f_\Omega(\Omega)\, A_{\rm eff}(\Omega, E_{\mathrm{fin}}|E_\Nu)\, .
\end{equation}Here $\Delta t$ is the total observation time; and $A_{\rm eff}$ includes the geometric area of the detector, detection efficiency, the absorption of the flux inside Earth, and the probability for the initial flux to produce a detectable signal with energy $E_{\rm fin}$. It encodes all the details of the propagation and absorption models, as we explain below.

Here, $(E_{\rm min},E_{\rm max})$ is a narrow energy window of the initial N flux, sufficient to explain the $\mathcal{O}({\rm EeV})$ ANITA-IV events. 
We assume $A_{\rm eff}$ does not depend strongly on energy within this narrow energy range~\cite{ANITA:2021xxh,IceCube:2016uab}, and then $A_{\rm eff}(\Omega, E_{\rm fin}|E_\Nu) = A_{\rm eff}(\Omega,E_{\mathrm{fin}})$, such that
\begin{equation}\label{eq:exp_events_0}
     \frac{\mathrm{d}N(\Omega)}{\mathrm{d}\Omega\,  \dd E_{\rm fin}} = 
     \left(\Phi_0\int_{E_{\text{min}}}^{E_\text{max}} \mathrm{d}E_\Nu\,  f_E(E_\Nu)\right) f_\Omega(\Omega) \,  \Delta t \, A_{\rm eff}(\Omega,E_{\mathrm{fin}})   \equiv 
     \Phi\, f_\Omega(\Omega)\, \Delta t \, A_{\rm eff}(\Omega,E_{\mathrm{fin}})
     \, .
\end{equation}
Then, for our KM3NeT and IceCube analyses we account for events only above 10 PeV, i.e., 
\begin{equation}\label{eq:exp_events}
     \frac{\mathrm{d}N(\Omega)}{\mathrm{d}\Omega} = 
     \Phi\, f_\Omega(\Omega)\, \Delta t \, \int_{\rm 10\, PeV}^{E_{\rm max}}\dd E_{\rm fin}\, A_{\rm eff}(\Omega,E_{\mathrm{fin}})
     \, .
\end{equation}

Finally, experiments do not perfectly reconstruct the true angle of the incoming particle, $\Omega^{\rm true}\equiv (\varphi^{\mathrm{true}},\theta^{\text{true}})$. To take this into account, we assume Gaussian angular uncertainty $(\Delta\varphi,\, \Delta\theta)$~\cite{ANITA:2020gmv,KM3NeT:2025npi}. 
Then, the expected number of events per unit solid angle as a function of the reconstructed angle $(\varphi^{\mathrm{rec}},\,\theta^{\text{rec}})$ is given by
\begin{equation}
    \frac{\mathrm{d}\bar{N}(\Omega^{\text{rec}})}{\mathrm{d}\Omega} = 
	\int \mathrm{d}\Omega^{\text{true}} \frac{1}{2\pi\Delta\theta\,\Delta\varphi} 
    \exp\left[ -\frac{(\theta^{\text{rec}}-\theta^{\text{true}})^2}{2(\Delta\theta)^2} -\frac{(\varphi^{\rm rec}-\varphi^{\rm true})^2}{2(\Delta \varphi)^2} \right] 
    \frac{\mathrm{d}N(\Omega^{\text{true}})}{\mathrm{d}\Omega} 
\end{equation}

%%%%%%%%%%%%%%%%%%%%%%%%%%%%%%%%%%%%%
\subsection{Absorption and detection processes}
\label{sec:probabilities}
%%%%%%%%%%%%%%%%%%%%%%%%%%%%%%%%%%%%%
As described in the main text, there are four possible event sources: N scattering, T decay, T scattering, and $\mu$ tracks. The effective area for T decay is
\begin{equation}
    A_{\rm eff, dec}^\Tau = P_{\rm exit}^\Tau P_{\rm decay}^\Tau A_g\, \varepsilon\, ,
\end{equation}
where $P_{\rm decay}^\Tau = 1-e^{-d/c\tau}$ is the probability that T decays inside the instrumented volume, $A_g$ is the geometrical area and $\varepsilon$ the detector efficiency. $P_{\rm exit}^\Tau$ is the probability that an N particle creates a T particle and the latter arrives to the detector. For a medium of length $L$ and constant mean free path $\lambda = (n\sigma)^{-1}$, with $n$ the nucleon number density of the medium, this is given by
\begin{equation}
    P_{\rm exit}^\Tau = \frac{c\tau}{\lambda}e^{-L/\lambda}\left(1-e^{-L/c\tau}\right)\, ,
\end{equation}
where we have assumed that T can be absorbed by matter with cross-section $\sigma$. This probability also enters in the effective area to T scatterings,
\begin{equation}
    A_{\rm eff,\, scat}^\Tau = P^\Tau_{\rm exit}\, \sigma N_{\rm targets}\, \varepsilon\, ,
\end{equation}
with $N_{\rm targets}$ the number of targets in the instrumented volume of the detector.

Scatterings can also be produced by the interaction of N particles, that is,
\begin{equation}
    A_{\rm eff,\, scat}^\Nu = P^\Nu_{\rm exit}\, \sigma N_{\rm targets}\, \varepsilon\, ,
\end{equation}
where $P^\Nu_{\rm exit}$ is the probability that an N particle arrives to the detector. For a medium of constant density,
\begin{equation}
    P^\Nu_{\rm exit} = e^{-L/\lambda}\left[1+
    \left(\frac{c\tau}{\lambda}\right)^2\left(\frac{L}{c\tau}+e^{-L/c\tau}-1\right)\right]\, .
\end{equation}
Here we assume that N arrives to the detector because it has not been absorbed yet, or because the secondary T has been absorbed by Earth and produced a tertiary N. We find that a single regeneration process is sufficient for the required level of precision of our analysis.

Finally, the effective area to muon tracks is given by
\begin{equation}\label{eq:muon-effective-area}
    A_{\rm eff}^\mu = A_g\,\varepsilon \int_{\mathrm{10\, PeV}}^{E_{\rm max}}\dd E_{\rm fin}^\mu P_\mu(E^\mu_{\rm fin})\, ,
\end{equation}
where $P_\mu(E^\mu_{\rm fin})$ is the probability that a muon with energy $E_{\rm fin}^\mu$ arrives to the detector, accounting for its energy losses. These are described in the next section.

Now the effective areas to cascades, starting tracks and through-going muons can be defined as
\begin{equation}\label{eqa:effective-areas-topos}
    \begin{split}
        A_{\rm eff}^{\rm casc} &= A_{\rm eff,dec}^\Tau \left[1-\mathrm{Br}(\Tau\to\mu)\right] + A_{\rm eff,scat}^\Tau + A_{\rm eff}^\Nu\, , \\
        A_{\rm eff}^{\rm start} &= A_{\rm eff,dec}^\Tau\, \mathrm{Br}(\Tau\to\mu)\, , \\
        A_{\rm eff}^{\rm through} &= A_{\rm eff}^\mu\, \mathrm{Br}(\Tau\to\mu)\, ,
    \end{split}
\end{equation}
respectively. The total effective area is $A_{\rm eff}^{\rm tot} = A_{\rm eff}^{\rm casc}+A_{\rm eff}^{\rm start}+A_{\rm eff}^{\rm through}$.

%%%%%%%%%%%%%%%%%%%%%%%%%%%%%%%%%%%%%
\subsection{Muon energy losses}\label{sec:muon-energy-losses}
%%%%%%%%%%%%%%%%%%%%%%%%%%%%%%%%%%%%%

While muon lose energies by stochastic processes, they can be modeled in a continuous manner, i.e., by taking $\dd E_\mu/\dd x = \langle  \dd E_\mu/\dd x \rangle$, where $x$ is the column depth. In this ``continuous-slowing-down-approximation'', the loss of energy is usually parameterized as~\cite{ParticleDataGroup:2024cfk}
\begin{equation}
- \frac{\dd E_\mu}{\dd x} = a(E_\mu)+b(E_\mu)E_\mu\, ,
\end{equation}
where $a(E)$ is the electronic stopping power and $b(E)$ is due to radiative processes ionization, bremsstrahlung, photonuclear interactions, and electron pair production.
Both $a(E)$ and $b(E)$ are slowly-varying functions. Radiative losses dominate above the so-called critical energy $E_{\mu c}$, defined by $a(E_{\mu c}) = E_{\mu c}\, b(E_{\mu c})$ for muons with $E_{\mu c}\sim 200\, \mathrm{GeV}$. Then, it is a good approximation that,
\begin{equation}
- \frac{\dd E_\mu}{\dd x} = b(E_\mu)E_\mu\, .
\end{equation}
For constant $b(E_{\mu c})$, this gives exponential energy losses, such that the distance travelled by a muon between energy $E_{\rm ini}$ and $E_{\rm fin}$ is
\begin{equation}
    d_\mu = \xi\, \log\frac{E_{\rm ini}+E_{\mu c}}{E_{\rm fin}+E_{\mu c}}\, ,
\end{equation}
where $\xi$ is the characteristic length scale for propagation of high-energy muons. Then, in this approximation, all muons that are created with energy $E_{\rm ini}$ at a distance less than $d_\mu$ from the detector will arrive to the detector with energy larger than $E_{\rm fin}$. In the SM, this gives an effective volume to muon neutrinos $V = A_g\,d_\mu$, with $A_g$ the geometric area of the detector. 

While tabulated values of $b(E_\mu)$ exist~\cite{ParticleDataGroup:2024cfk}, the distance $d_\mu$ traveled by UHE muons is better estimated using Monte Carlo generated data.
In this work, we have propagated muons through water for distances up to $34\,$km (the distance from ARCA to the cliff), using the Muon Monte Carlo (MMC) code~\cite{Chirkin:2004hz}. For distances, $r$, greater than $34\, \rm km$ muons were propagated first through $(r - 34)\rm\, km$ of rock, and then $34\rm \, km$ of water. We have checked that $d_\mu$ computed in this manner matches tabulated values, and thus we use the ansatz in~\cite{GaisserEngelResconi}. 

\begin{figure}
  \centering
    \includegraphics[width=0.495\linewidth]{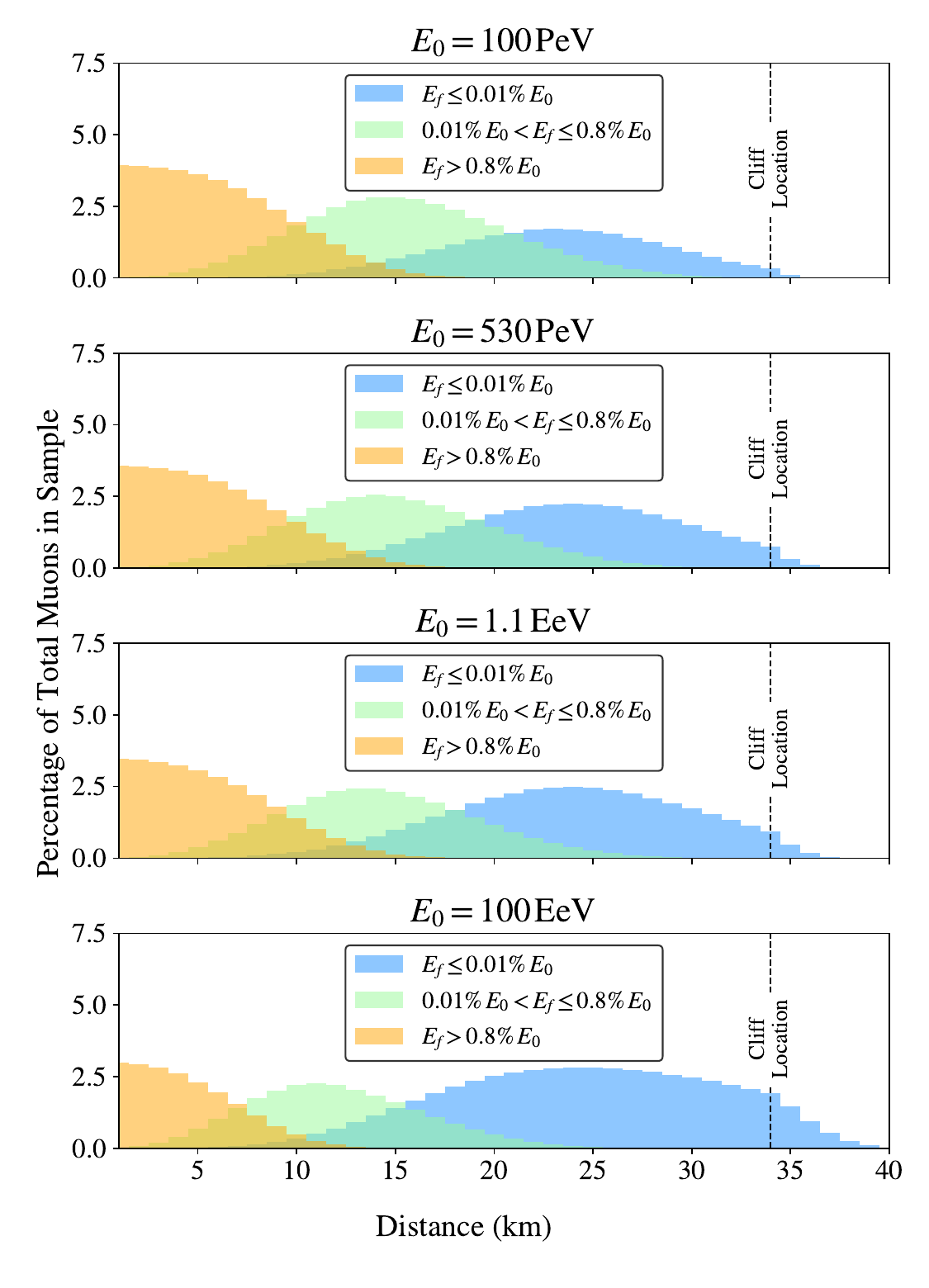}
    \includegraphics[width=0.495\linewidth]{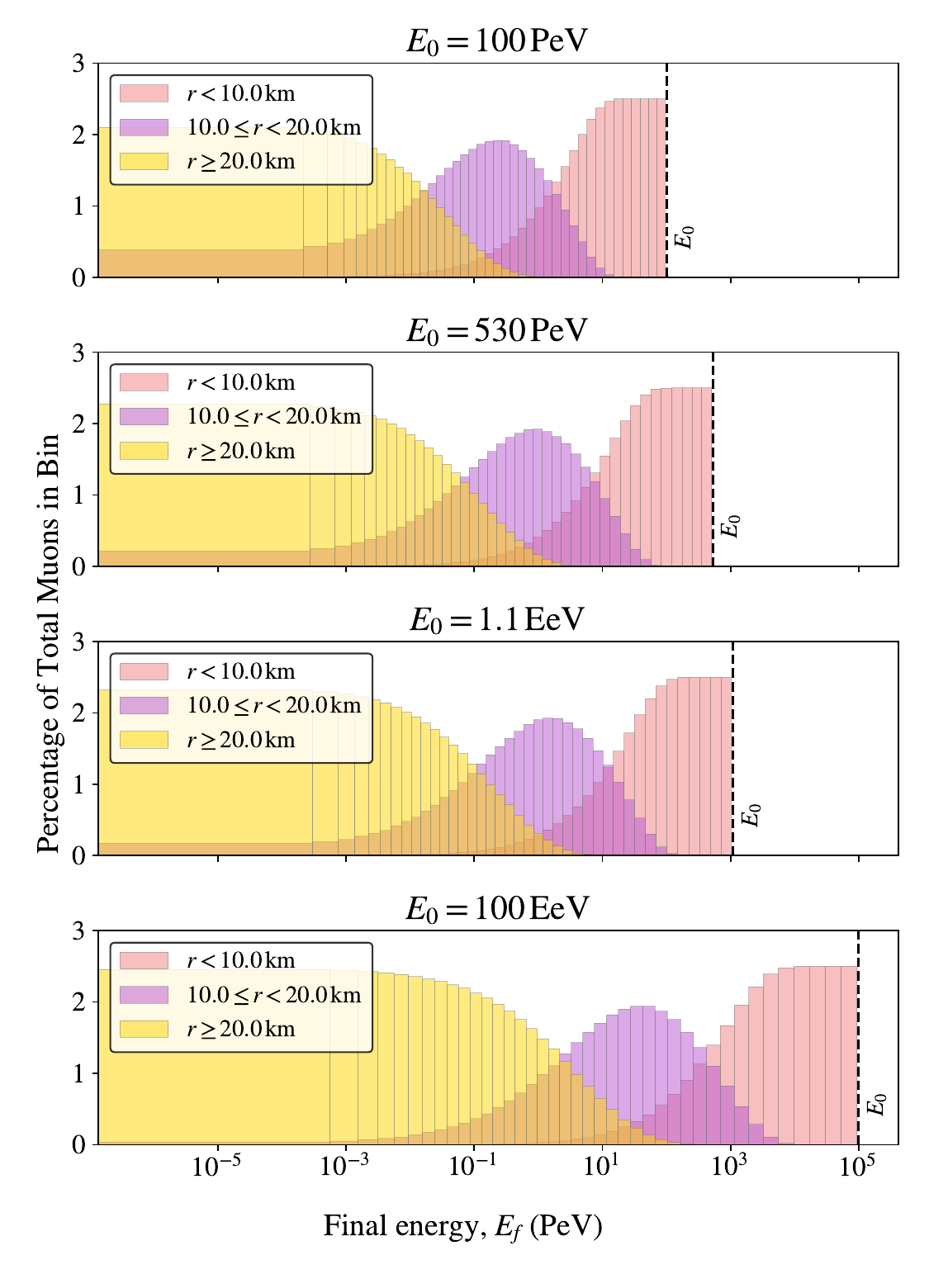}
  \caption{Distance and final-energy histograms for four different initial energies, \(E_0\), for muons propagated through water up to 34 km and rock beyond 34 km. Left: Percentage of muons surviving through some distance for the three final energy, \(E_f\), slices. 
  The height of each bin corresponds to the percentage of the total number of muons in the sample, including all distances and final energies. Right: Percentage of muons with some final energy for the three distance, \(r\), ranges. 
  Bar heights show the percentage of the total muons in each bin, further broken down by distance slices indicated by color. Heights differ across bins within a given distance slice, but the sum of counts in each distance range in each bin is approximately constant across bins.}
  \label{fig:equalcount_distance_grid}
  \label{fig:equalcount_Ef_grid}
  \label{fig:equalcount_sidebyside}
\end{figure}

\Cref{fig:equalcount_Ef_grid} (left) shows overlaid distance histograms with three ranges of final energy and traveled distance, for four different initial energies. The cuts in \(E_f\) are proportional to each respective injected energy, and the proportions are the same for each plot. The height of the bins corresponds to the percentage of muons in the total sample shown in the plot. The location of the cliff indicated on the plot corresponds to the distance from the ARCA detector to the cliff for the most probable particle trajectory~\cite{KM3NeT:2025npi}. For the direction of the event, it is highly unlikely that a muon with an initial energy on the order of hundreds of PeV could propagate through greater than 34 km of water. Similarly, \cref{fig:equalcount_distance_grid} (right) shows overlaid final energy histograms with three ranges of distance for four different initial energies. Distance ranges here are the same for each plot. 

\begin{figure}
    \centering
    \includegraphics[width=0.54\linewidth]{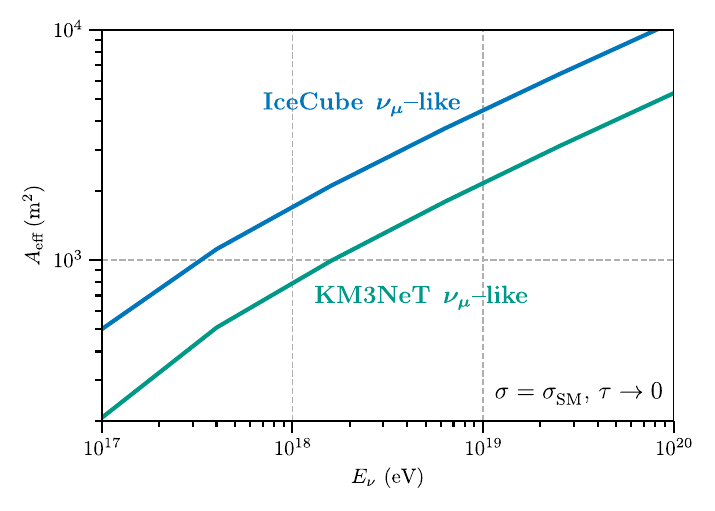}
    \caption{Total effective areas for IceCube and KM3NeT in the $\nu_\mu$-like scenario, all-sky averaged, as defined in \cref{eqa:effective-areas-topos}.}
    \label{fig:effective-areas}
\end{figure}

Then, the probability for a muon being produced from decay at a distance $x$ from the detector is given by
\begin{equation}\label{eq:prob-muon-x}
    \frac{\dd P_\mu(x)}{\dd x}= \frac{1}{\lambda}e^{-(d-x)/\lambda}\left(1-e^{-(d-x)/c\tau}\right)\, ,
\end{equation}
for a homogeneous medium of depth $d$.
Then, for a given $E_{\rm max}$, the integral over $E_{\rm fin}$ at \cref{eq:muon-effective-area} can be understood as an integral over distance $x$, 
\begin{equation}
    A_{\rm eff}^\mu = 
    A_g\, \varepsilon\, \int_{0}^{d_\mu(10\, {\rm PeV})}\dd x\, \frac{\dd P_\mu(x)}{\dd x} =
    A_g\, \varepsilon\left[
        e^{-d/\lambda}\left(e^{d_\mu/\lambda}-1\right)-
        \frac{c\tau}{c\tau+\lambda}e^{-d\left(\frac{1}{c\tau}+\frac{1}{\lambda}\right)}
        \left(e^{d_\mu\left(\frac{1}{c\tau}+\frac{1}{\lambda}\right)}-1\right)
    \right]\, .
\end{equation}
And, for any $d<d_\mu$, one must substitute $d_\mu \to d$. Probabilities for a medium with layers of different density can be found \href{https://github.com/tbertolez/BSMatUHEdets}{in our code~\faGithub}. \Cref{fig:effective-areas} shows the effective areas of IceCube and KM3NeT computed in this manner, in the $\nu_\mu$-like scenario, that is, 
\begin{equation}
    \sigma(E_\nu) = \sigma_{\rm SM}(E_\nu)\simeq
    2.35\times 10^{-32}\, \mathrm{cm}^2\left(\frac{E_\nu}{10\, \mathrm{EeV}}\right)^{0.363}
\end{equation}
and $\tau \to 0$. We have calibrated the KM3NeT geometric area, $A_g$, to match the official result from the collaboration in the $\nu_\mu$-like scenario~\cite{KM3NeT:2025npi}.

%%%%%%%%%%%%%%%%%%%%%%%%%%%%%%%%%%%%%
\subsection{Test statistics}
\label{sec:test-statistics}
%%%%%%%%%%%%%%%%%%%%%%%%%%%%%%%%%%%%%
In this section, we describe our statistical analysis. As data is scarce, an unbinned Poisson likelihood is well-suited. For each experiment, the test statistic is given, up to constants, by
\begin{equation}
    \mathcal{TS}^{(\rm exp)}(\Phi,\sigma,\tau,\rm Br) =
2\int \dd \Omega\,  \mu(\Omega;\Phi,\sigma,\tau,\rm Br)
- 2\sum_{i=1}^N\log\tilde{\mu}(\Omega^{\text{rec}}_i;\Phi,\sigma,\tau,\rm Br)  
\, .
\end{equation}
For each particular experiment, ${\mu \equiv \dd N/\dd \Omega}$, $\tilde{\mu}(\theta^{\text{rec}}) \equiv \sin\theta^{\text{rec}} \dd \bar{N}/\dd \Omega(\theta^{\text{rec}},\varphi^{\rm rec})$, and $(\varphi^{\rm rec}_i,\theta_i^\mathrm{rec})$ are the reconstructed angles of the observed events. Here, ${\rm Br}\equiv {\rm Br}(\Tau\to\mu)$.

For instance, the test statistic for KM3NeT is
\begin{equation}\label{eq:ts-km3net}
\begin{split}
    \mathcal{TS}^{(\rm K3)}(\Phi,\sigma,\tau,\rm Br) =\ &
    \frac{\Delta t^{\rm (K3)}\, \Phi}{2\pi}\int_{10 \, \rm PeV}^{E_{\rm max}} \dd E_{\rm fin}\int \dd \Omega\,  (A_{\rm eff}^{\rm start}+A_{\rm eff}^{\rm through})(E_{\rm fin},\Omega;\sigma,\tau,\rm Br) \\ 
    &- 2\log\left(\frac{\Delta t^{\rm (K3)}\Phi}{4\pi}\int_{10 \, \rm PeV}^{E_{\rm max}}\dd E_{\rm fin}\, \bar{A}_{\rm eff}^{\rm through}(E_{\rm fin},\Omega^{\rm rec};\sigma,\tau,\rm Br)\right)  
\, ,
\end{split}
\end{equation}
where
\begin{equation}
    \bar{A}_{\rm eff}^{\rm through}(\Omega^{\rm rec}) = 
    \sin\theta^{\rm rec}\int \mathrm{d}\Omega^{\text{true}} \frac{1}{2\pi\Delta\theta\,\Delta\varphi} 
    \exp\left[ -\frac{(\theta^{\text{rec}}-\theta^{\text{true}})^2}{2(\Delta\theta)^2} -\frac{(\varphi^{\rm rec}-\varphi^{\rm true})^2}{2(\Delta \varphi)^2} \right] 
    A^{\rm through}_{\rm eff}(\Omega^{\rm true})\, .
\end{equation}
The official analysis of the collaboration has only-looked for track-like signals produced by a muon~\cite{KM3NeT:2025npi}. Since possible cascade-like signals may have not been accounted for, these are not included in the test statistic as defined in the main text. If the KM3NeT analysis had looked for cascades, then the first term would also include $A_{\rm eff}^{\rm casc}$, and would be the integral over the total effective area $A_{\rm eff}^{\rm tot}$.

\begin{figure}
    \centering
    \includegraphics[width=0.53\linewidth]{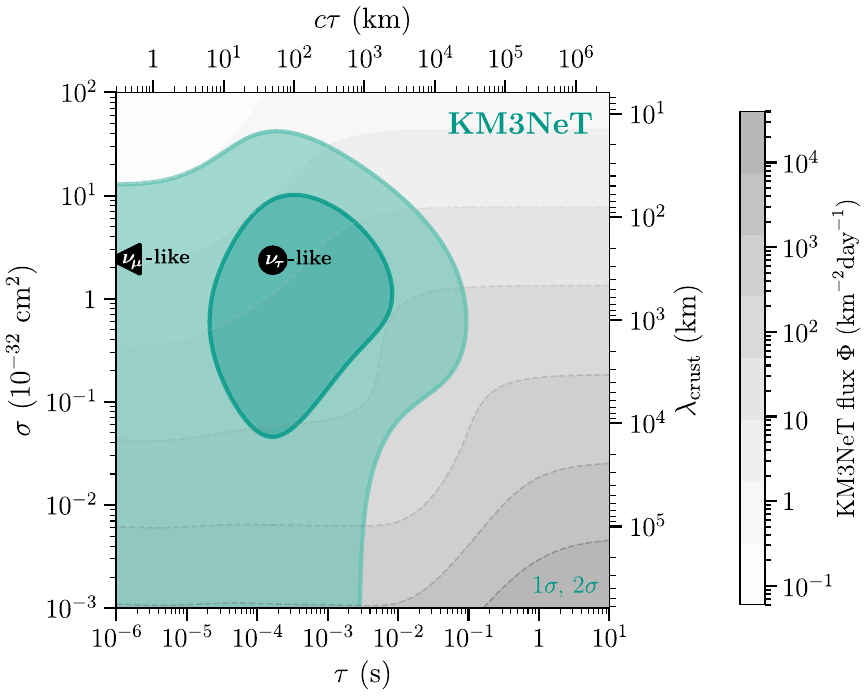}
    \caption{Analogous to~\cref{fig:KM3NeT-test-statistic}, but if KM3NeT analysis had looked for cascade-like events. Larger lifetimes are now more excluded, where track-like events are disfavored against cascade-like events.}
    \label{fig:test-statistic-cascades}
\end{figure}

For a given $(\sigma,\tau)$, the best fit to \cref{eq:ts-km3net} is always at $\rm Br = 1$ which optimizes the probability of producing a muon and 
\begin{equation}\label{eq:k3-only-bf-flux}
    \Phi_{\rm K3}^{\rm BF}(\sigma,\tau)= \frac{4\pi}{\Delta t^{\rm(K3)}\int\dd E_{\rm fin}\,\dd \Omega\,  (A_{\rm eff}^{\rm start}+A_{\rm eff}^{\rm through})}\, ,
\end{equation}
where the integral over $E_{\rm fin}$ is above 10 PeV. This flux normalization exactly fits the number of events measured (i.e., one). Profiling over $(\rm Br,\Phi)$, the KM3NeT-only test statistic is only constrained by the direction of KM3-230213A and the fact that this is a through-going muon. Quantitatively, this is
\begin{equation}
    \mathcal{TS}^{(\rm K3)}(\sigma,\tau) = 
    -2\log\left(\dfrac{\int \dd E_{\rm fin} \bar A_{\rm eff}^{\rm through}(\Omega^{\rm rec},\rm Br = 1)}{\int\dd E_{\rm fin}\,\dd \Omega\,  (A_{\rm eff}^{\rm start}+A_{\rm eff}^{\rm through})(\Omega,\rm Br = 1)}\right)\, .
\end{equation}
This is the test statistic shown in \cref{fig:KM3NeT-test-statistic}, with the flux in the background given by \cref{eq:k3-only-bf-flux}. If KM3NeT had looked for cascade-like signals, then we would replace $(A_{\rm eff}^{\rm start}+A_{\rm eff}^{\rm through})$ for $A_{\rm eff}^{\rm tot}$, which would result in the contours shown in \cref{fig:test-statistic-cascades}. This test statistic puts more stringent bounds to long lifetimes, where the production rate of through-going muons is smaller compared to the signals produced by the primary vertex.

In ANITA, muon tracks are not observable, in our class of BSM models all events are monochromatic and we assume axial symmetry. Then, its test statistic is analogous to the one defined in~\cite{Bertolez-Martinez:2023scp},
\begin{equation}\label{eq:ts-ANITA-diffuse-withphi} 
\begin{split}
 \mathcal{TS}^{(\textrm{AN})}(\Phi,\sigma,\tau,\rm Br) = & 
 -2\sum_{i=1}^4\log\left(\frac{\Phi\,\Delta t^{\rm (AN)}}{4\pi} \bar{A}_{\rm eff}^{\rm casc}(\theta^{\text{rec}}_i;\sigma,\tau,\rm Br)\right) + \Phi\,\Delta t^{\rm(AN)}\int \mathrm{d}\theta\, \sin\theta\, A_{\rm eff}^{\rm casc}(\theta;\sigma,\tau,\rm Br)\, .
\end{split}
\end{equation}

The test statistic for the KM3NeT-ANITA joint fit is then $\mathcal{TS}^{\rm (A3)}\equiv \mathcal{TS}^{(\rm K3)}+\mathcal{TS}^{(\rm AN)}$. For given $(\sigma,\tau,\rm Br)$, the best-fit flux is
\begin{equation}\label{eq:a3-phi-bf}
    \Phi_{\rm BF}^{\rm A3}(\sigma,\tau,{\rm Br}) = 
    \frac{4\pi(n^{\mathrm{K3}}+n^{\mathrm{AN}})}{\Delta t^{\rm(K3)}\int\dd E_{\rm fin}\,\dd \Omega\,  (A_{\rm eff,K3}^{\rm start}+A_{\rm eff,K3}^{\rm through})+2\pi\Delta t^{(\rm AN)}\int \sin\theta\, \dd\theta\, A_{\rm eff, AN}^{\rm casc}}
\end{equation}
where $n^{\mathrm{K3}}=1,\,n^{\mathrm{AN}}=4$ are the total number of events observed at KM3NeT and ANITA-IV, respectively. For this best-fit flux, the best fit for the branching ratio is found at
\begin{equation}\label{eq:a3-br-bf}
    \mathrm{Br}_{\rm BF}^{\rm A3} = 
    \left(1+\frac{n^{\mathrm{AN}}}{n^{\mathrm{K3}}}
    \frac{\Delta t^{\rm (K3)}\int \dd E_{\rm fin}\, \dd\Omega\, (A_{\rm eff,dec,K3}^{\rm T}+A_{\rm eff,K3}^\mu)}
    {2\pi\Delta t^{(\rm AN)}\int\sin\theta\, \dd \theta\, A_{\rm eff,AN}^{\rm casc}}
    \right)^{-1}\, .
\end{equation}
Then, \cref{fig:ANITAK3-test-statistic} draws the contours of $\mathcal{TS}^{(\rm A3)}$ for this best-fit flux and branching ratio.

Finally, since IceCube has not seen any event, its test statistic is given by
\begin{equation}\label{eq:ts-IC-diffuse-withphi} 
 \mathcal{TS}^{(\textrm{IC})}(\Phi,\sigma,\tau,{\rm Br}) =  
  \Phi\,\Delta t^{\rm(IC)}\int \mathrm{d}\theta\, \sin\theta\, A_{\rm eff}^{\rm tot}(\theta;\sigma,\tau,{\rm Br})\, .
\end{equation}
The total best fit is $\mathcal{TS}^{(\rm AI3)}=\mathcal{TS}^{(\textrm{K3})}+\mathcal{TS}^{(\textrm{AN})}+\mathcal{TS}^{(\textrm{IC})}$. The best-fit flux and branching ratio are still analytic and can be found in \href{https://github.com/tbertolez/BSMatUHEdets}{our code~\faGithub}.

\subsection{Spectrum of muons at $E_\mu<10\, \rm PeV$}\label{sec:subPeV-muons}
\begin{figure}
    \centering
    \includegraphics[width=0.495\linewidth]{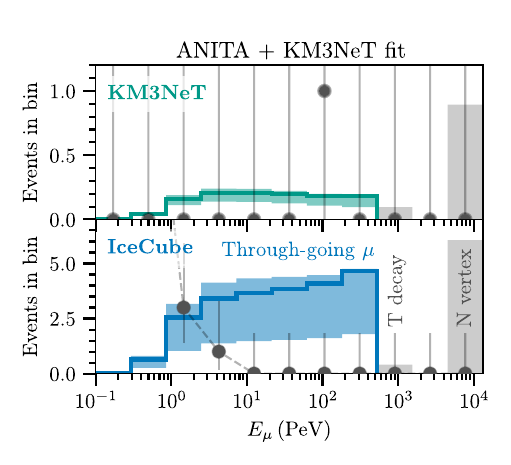}
    \includegraphics[width=0.495\linewidth]{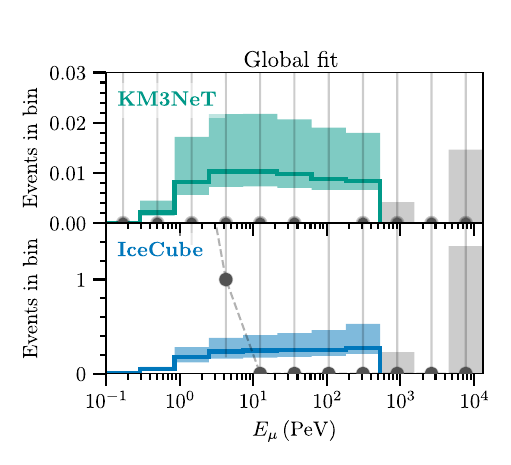}
    \caption{Energy distribution of the expected events for the ANITA+KM3NeT best fit (left panel) and the global best fit including also IceCube (right panel). Teal (blue) curves show the spectrum of through-going muons due to energy losses from their production to their detection at ARCA (IceCube), in the best fit of the corresponding fit. Shaded regions show the one sigma theoretical uncertainty. The gray bar at $E\sim 5\, \rm EeV$ accounts for signals from the scattering of the primary particle, N, while the gray bar at $E\sim 1\, \rm EeV$ accounts for signals from the decay of the secondary particle, T, inside the detector. Black data points show the measured through-going muons at the detectors~\cite{KM3NeT:2025npi,Abbasi:2021qfz}, with a gray dashed line as a visual guide to IceCube high-energy events.}
    \label{fig:muon-tail}
\end{figure}
In the main text, we have considered only muons above $10\, \rm PeV$, as in the official collaboration analysis. However, IceCube and KM3NeT have measured through-going muons at lower energies. In our framework, while we are injecting a flux only at energies $\mathcal{O}(\mathrm{EeV})$, we can still have muons with arbitrarily low energies if they travel long enough. In this section, we check that incorporating $E_\mu<10\, \rm PeV$ muons does not modify our conclusions. 

For a given injected energy, the MMC data shown in \cref{fig:equalcount_Ef_grid} is a probability density for the distance traveled by the muon $x$ and the final energy of the muon $E_{\rm fin}$, $f(x,E_{\rm fin})$. Then, for a given final energy bin $(E_i,E_{i+1})$, the effective area to through-going muons is
\begin{equation}
    A_{\rm eff}^{\mu,i} = A_g\int_{E_i}^{E_{i+1}}\dd E_{\rm fin}\, \varepsilon(E_{\rm fin})\int_0^{d}\dd x \, \frac{\dd P_\mu(x)}{\dd x}\,f(x,E_{\rm fin})\, ,
\end{equation}
with $\dd P_\mu(x)/\dd x$ given in \cref{eq:prob-muon-x}, $d$ the length of the trajectory inside Earth, and $\varepsilon(E_{\rm fin})$ the detector efficiency, which decreases to lower energies~\cite{Palomares-Ruiz:2015mka}. Using this energy-dependent effective area, one can compute the number of through-going muons at PeV and sub-PeV energies.

On the one hand, the left panel of \cref{fig:muon-tail} shows the full energy spectrum of muons at KM3NeT and IceCube, for the best-fit $(\sigma,\tau,\Phi,{\rm Br})$ in the ANITA+KM3NeT analysis of \cref{sec:anita-iv}. As shown in \cref{sec:icecube}, these experiments are in tension with IceCube for $E_{\rm fin}>10\, \rm PeV$. This plot shows that the main source of tension with IceCube is muons with $E_{\rm fin}>10\, \rm PeV$, and that considering lower-energy muons does not strongly modify the tension.

On the other hand, the right panel of \cref{fig:muon-tail} shows the best fit of the global analysis, including IceCube. As explained in \cref{sec:icecube}, including IceCube reduces the flux normalization and the predicted number of events. The tension still remains, but is reduced. 
The figure shows that extending the analysis to $E_{\rm fin}<10\, \rm PeV$ does not add further tension to the analysis, and the conclusions reached in the main text remain valid.

%%%%%%%%%%%%%%%%%%%%%%%%%%%%%%%%%%%%%%%%%%%%%%%%%%%%%%%%%%%%%%%%%%%%%%%%%%%%%%
\section{Results under different assumptions}
\label{sec:absorption}
%%%%%%%%%%%%%%%%%%%%%%%%%%%%%%%%%%%%%%%%%%%%%%%%%%%%%%%%%%%%%%%%%%%%%%%%%%%%%%

The analysis presented in the main text contains different assumptions on both the particle physics effective model and the experimental exposures, which we have enumerated in \Cref{sec:formalism}. In this Appendix, we further discuss these choices, quantify their impact on the final results, and show that the conclusions drawn in \Cref{sec:conclusions} remain robust even if these assumptions are relaxed.

\subsection{Energy transfer to muons}\label{app:muon-transfer}
In the simplified model-independent approach considered in the main text, we have taken the simplifying assumption that, if a T particle with energy $E_\Tau$ decays into a muon, this muon is always produced with the same energy $E_{\rm ini}$. In other words, that $E_{\rm ini} = y_{\Tau\to\mu}E_\Tau$ with $0<y_{\Tau\to\mu}<1$ a fixed parameter. However, the energy spectrum of muons after decaying will depend on the particular particle physics model. For instance, in the SM $y_{\tau\to\mu}$ is well distributed within 0.05 and 1~\cite{Cummings:2019rlt}. In our model, the choice of $y_{\Tau\to\mu}$ determines how many muons can arrive to the detector with $E_{\rm fin}>10\,\rm PeV$, i.e., the muon exposure of KM3NeT and IceCube.

\begin{figure}[b]
    \centering
    \includegraphics[width=0.495\linewidth]{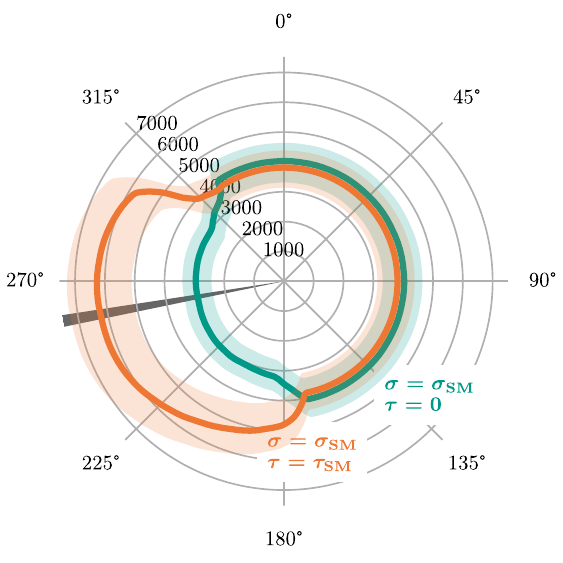}
    \caption{Analogous to~\cref{fig:azimuthal-plot}. The solid line corresponds to $y_{\Tau\to\mu} = 0.5$ (as in the main text), while the colored bands correspond to $0.25 \leq y_{\Tau\to\mu}\leq 0.9$.} 
    \label{fig:azimuthal-plot-different-ys}
\end{figure}

The analysis reported in the main text chooses a minimal agnostic assumption, $y_{\Tau\to\mu} = 0.5$. \Cref{fig:azimuthal-plot-different-ys} shows how larger $y_{\Tau\to \mu}$ increase the effective area. In particular, varying between $y_{\Tau\to\mu} = 0.9$ (outer ring) and $y_{\Tau\to\mu}=0.25$ (inner ring) varies the effective area by 20\% with respect to the central value. 

Indeed, we have checked that the effect of $y_{\Tau\to\mu}$ matches in a very good approximation an overall scaling of $A_{\rm eff}^\mu$. 
Since the statistical tests defined in \cref{sec:test-statistics} depend on ratios of effective areas, the effect of varying $y_{\Tau\to\mu}$ in this range on the $(\sigma,\tau)$ preferred regions of \cref{fig:KM3NeT-test-statistic} is negligible. However, the flux normalization required to produce the KM3NeT event varies within $\Phi \simeq (19-25)\, \rm km^{-2}\, day^{-1}$. In \Cref{app:exp-variation} we quantify the impact of this effect in the joint fits. 

\subsection{Dependence of ANITA-IV exposure on decay altitude}\label{app:decay-altitude}

In the analysis presented in the main text, we have used a simplified likelihood for the ANITA-IV experiment. As computed in~\cite{PierreAuger:2025hvl}, the real ANITA exposure depends on the altitude of the T decay point, and on the energy deposited in the shower. In this section, we estimate the impact of the dependence on decay altitude in our results.

In our analysis, T can decay and produce the shower at any point before $d$ (the distance between ANITA-IV and the Earth's exit point), and we assume that at any point the radio signal from the shower will be equally visible. With this assumption, the probability of a decay being detected only depends on the decay probability, $P_{\rm decay}^\Tau = 1-e^{-d/c\tau}$, and the efficiency $P_{\rm trig}$ from Fig. 10 of~\cite{ANITA:2021xxh} (which is provided in the region of maximal sensitivity). These probabilities do not depend on the altitude of the decay, $h$.

In reality, the ANITA exposure decreases with the altitude of the T decay point, as shown in Figure 3 of~\cite{PierreAuger:2025hvl}. However, the exposure provided in this figure corresponds to trajectories much below the horizon, far away from the direction of ANITA maximal sensitivity (and the ANITA-IV events), and cannot be directly implemented in our analysis. As a rough estimate, we take $\varepsilon(h_i,1\, \rm EeV)$ from Fig. 3 of~\cite{PierreAuger:2025hvl} and define
\begin{equation}\label{eq:altitude-decay}
    P_{\rm decay}^\Tau(\tau,\theta)\bar P_{\rm trig} = \sum_{i} w(h_i)\left(e^{-x(h_i)/c\tau}-e^{-x(h_{i+1})/c\tau}\right)\, ,\qquad \mathrm{with}\ w(h_i) = \xi(P_{\rm trig})\frac{\varepsilon(h_i,1\, \rm EeV)}{\sum_i\varepsilon(h_i,1\, \rm EeV)}\, .
\end{equation}
Here, the exponentials describe the probability of decaying within the layer $(h_i,h_{i+1})$, with $x(h_i)$ the projection of $h_i$ along the trajectory of T. Then, $w(h_i)$ encode the probability of a decay at height $h_i$ being detected. These weights follow the shape from Fig. 3 of~\cite{PierreAuger:2025hvl}, but are normalized such that we retrieve $P_{\rm decay}^\Tau(\tau,\theta) P_{\rm trig}$ from Fig. 10 of~\cite{ANITA:2021xxh}, for the SM values $\sigma = \sigma_{\rm SM},\, \tau = \tau_{\rm SM}$.

\begin{figure}[b]
    \centering
    \includegraphics[width=0.48\linewidth]{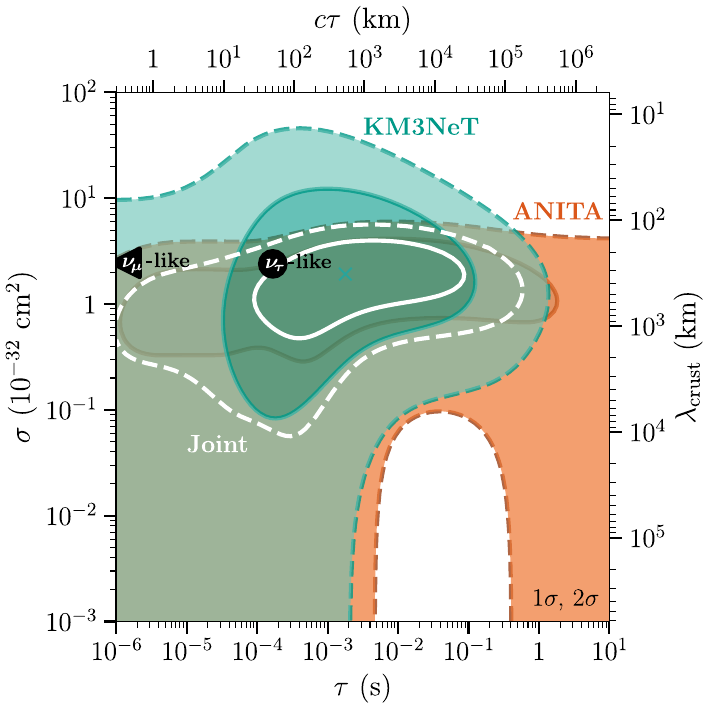}
    \caption{Analogous to~\cref{fig:ANITAK3-test-statistic}, but including the dependence of ANITA-IV exposure on the altitude of the decay point as explained in \cref{eq:altitude-decay}. 
    The ANITA-IV one sigma region is displaced to shorter lifetimes and the joint allowed region shows not significant change compared to \cref{fig:ANITAK3-test-statistic}.}
    \label{fig:test-statistic-altitude}
\end{figure}
\Cref{fig:test-statistic-altitude} shows that, with this correction, the one sigma region for ANITA-IV shifts slightly to shorter lifetimes. This is compatible with ANITA exposure peaking at low altitudes. As explained in the main text, the freedom on $\Phi$ and $\rm Br(\Tau\to \mu)$ can compensate for these changes, as shown in \cref{eq:a3-phi-bf,eq:a3-br-bf}. Thus, the ANITA+KM3NeT compatibility remains intact. This is not the case for the ANITA+KM3NeT+IceCube joint fit, which is affected by the best-fit lifetime decreasing to $\sim 0.6$ s, and has its tension increased to 3.2 sigma. 

Then, the refinement of the ANITA-IV likelihood indeed leads to quantitative differences. However, the qualitative conclusions 
drawn in the main text are robust, as we discuss in the next section. Also, note that the exposure in the angular range relevant to this analysis is not publicly available to our knowledge, so we could be missing systematics that compensate for the ones presented here. Thus, in the main text we choose to keep the simplified analysis.

Finally, the shower energy of the different ANITA-IV events varies between 0.8 and 3.6 EeV. Figure 3 of~\cite{PierreAuger:2025hvl} shows that ANITA exposure varies by 30\% within this energy range. We move now to show the possible effects of not considering this variation.

\subsection{Joint fits under variation of the assumptions} \label{app:exp-variation}
As we have seen in the previous two subsections, different choices of our assumptions and approximations can vary the total exposure of the different experiments, by up to 30\%. In this section, we redo the joint fits presented in the main text by varying the experimental exposures and show that our main conclusions discussed therein remain valid. In particular, we report the results of varying $0.25\le y_{\rm T\to\mu}\le 0.9$ and conservatively varying ANITA's exposure by a factor of 10, i.e., 
from $A_{\rm tot}/3$ to $3A_{\rm tot}$. 

In the ANITA+KM3NeT joint fit, the freedom on $\Phi$ and $\rm Br(T\to\mu)$ can compensate for any variation, as shown in \cref{eq:a3-phi-bf,eq:a3-br-bf}. Thus, the best-fit and allowed regions in the $(\sigma,\tau)$ parameter space remain the same. The flux normalization and branching ratio vary within
\begin{equation}
\begin{split}
    \Phi&\simeq (38-62)\, \rm km^{-2}\, day^{-1}\, ,\\
    \rm Br(T\to\mu)&\simeq (0.56-0.92)\, .
\end{split}
\end{equation}
Even if the preferred values for these two free parameters can vary, the ANITA-IV and KM3NeT
observations remain highly compatible and there are no significant modifications on the $(\sigma,\tau)$ preferred parameter space.

This statement is no longer true in the global fit. The fact that IceCube has seen no events introduces a $\Phi$-dependent likelihood. Thus, varying ANITA's exposure and $y_{\rm T\to\mu}$ in the range described above moves the best fit within the following intervals
\begin{equation}
\begin{split}
    \sigma &\simeq (0.6-1.0)\times 10^{-32}\, \rm cm^{2}\, ,\\
    \tau &\simeq (0.9-1.1)\times 10^{-3}\, \rm s\, ,\\
    \Phi&\simeq (1-3)\, \rm km^{-2}\, day^{-1}\, ,\\
    \rm Br(T\to\mu)&\simeq (0.15-0.32)\, .
\end{split}
\end{equation}
Note that the largest range of variation corresponds to the flux, as expected.

On the one hand, in the most optimistic side of these intervals--i.e.,  $y_{\Tau \to\mu} = 0.9$ and we have underestimated ANITA's exposure--, this predicts $\sim 0.014$ muons in KM3NeT, 2.89 events in ANITA-IV, and 0.71 events in IceCube. When confronted with observations, this scenario is excluded at 2.3 sigma. On the other hand, in the most pessimistic side, we expect $\sim 0.013$ muons in KM3NeT, 1.23 cascades in ANITA-IV, and 1.74 events in IceCube. This is a 3.3 sigma tension, of the same magnitude as that found in \Cref{app:decay-altitude} by adding the dependence with $h$.

In conclusion, while the uncertainties induced from our assumptions affect the numerical results as reported in this appendix, they do not effectively affect the conclusions reached in the main text. That is, that the KM3NeT and ANITA-IV observations are compatible, that their tension with IceCube can be slightly ameliorated with respect to the SM, and that under a diffuse flux hypothesis this tension remains above 2.3 sigma even considering the freedom of our model independent framework and the favorable hypothesis of a monochromatic flux. In particular, relaxing some of our main assumptions can worse the tension at the level of 3.3 sigma.

\subsection{T absorption}
\begin{figure}[b]
    \centering
    \raisebox{-\height}{\includegraphics[width=0.495\linewidth]{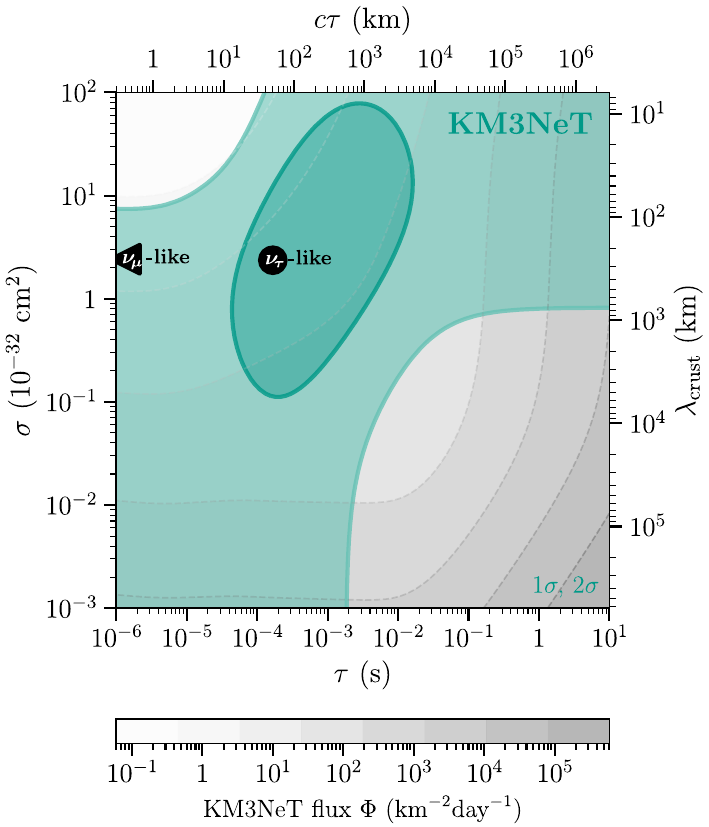}}
    \raisebox{-\height}{\includegraphics[width=0.495\linewidth]{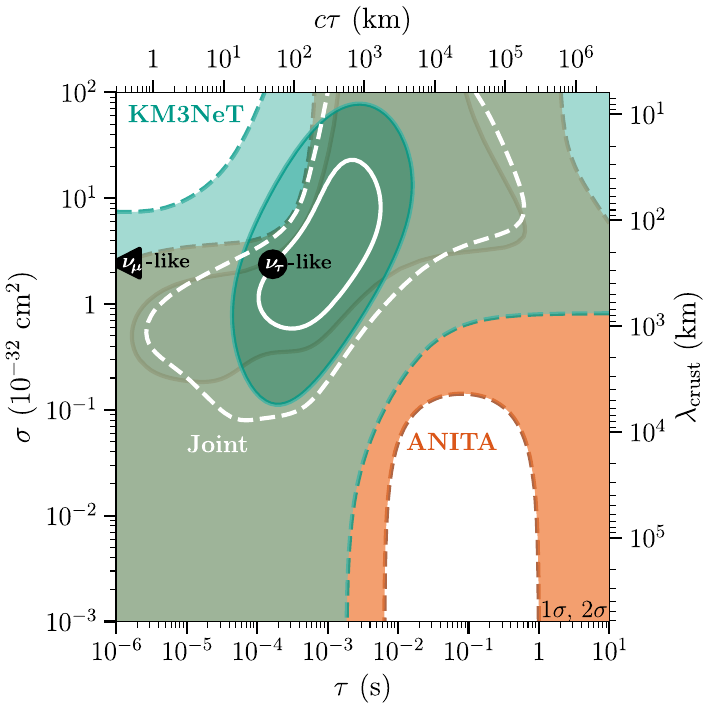}}
    \caption{Analogous to~\cref{fig:KM3NeT-test-statistic,fig:ANITAK3-test-statistic}, but considering that T cannot be absorbed by matter during its propagation. Values of $\sigma$ larger than in the case in which T can be absorbed are allowed .}
    \label{fig:no-absorption-test-statistiscs}
\end{figure}

\begin{figure}
    \centering
    \includegraphics[width=0.85\linewidth]{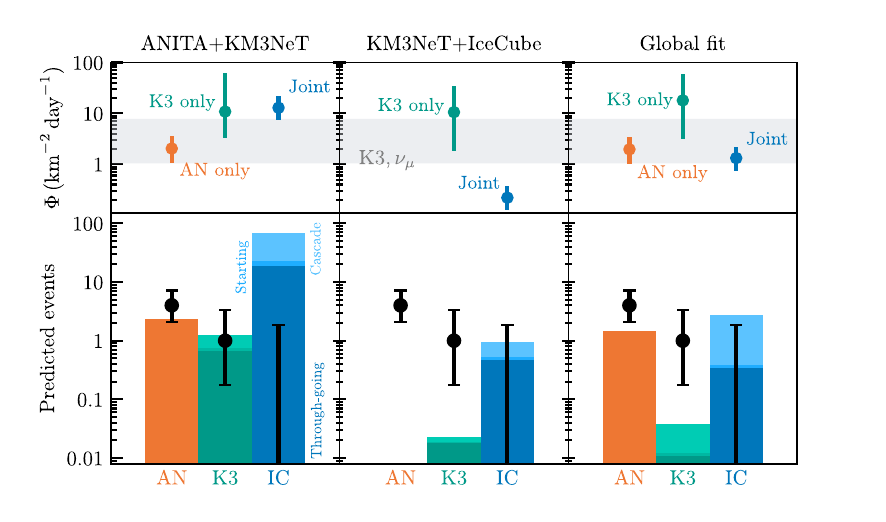}
    \caption{Analogous to~\cref{fig:flux-events}, but considering that T cannot be absorbed by matter during its propagation. This lowers the required flux normalizations, but do not have a great impact in the expected number of events at experiments.}
    \label{fig:no-absorption-flux-events}
\end{figure}

In the main text we have assumed time reversal symmetry and, thus, that T can be absorbed by the Earth with the same cross-section that produced it, $\sigma$. While this is expected in simple models, in general the N cross-section $\sigma_\Nu$ and the T cross-section $\sigma_\Tau$ do not need to be the same. $\sigma_\Tau=0$ increases the expected events in ANITA-IV and the produced muons. \Cref{fig:no-absorption-test-statistiscs} shows the allowed parameter space 
when $\sigma_\Tau =0$ and $\sigma_\Nu = \sigma$ for the KM3NeT (left panel) and KM3NeT+ANITA (right panel) analyses, respectively. The angular distribution of the events now allows large cross-sections, since the right lifetime $10^{-4}\, \mathrm{s}\lesssim \tau\lesssim 10^{-2}\, \rm s$ adjusts the right chord length from the observed trajectories. While the allowed region looks qualitatively different, \cref{fig:no-absorption-flux-events} shows that the predicted number of events for the different experiments is not significantly modified (even if fluxes are now smaller). Therefore, a similar tension among the different sets of data is still present and the conclusions drawn in the main text remain valid for $0\leq \sigma_\Tau\leq \sigma$.


\end{document}